\documentclass[twocolumn, twocolappendix]{aastex701}

\usepackage{CJK}
\usepackage{amsmath}	
\usepackage{amssymb}	
\usepackage{enumitem}
\usepackage{verbatim} 
\usepackage[caption=false]{subfig}
\usepackage{array}

\usepackage{listings}
\usepackage{hyperref}

\begin{document}
\begin{CJK*}{UTF8}{gbsn}

\title{The Persistent Thermal Anomalies in Rocky Worlds}


\newcommand{\todo}[1]{\textcolor{red}{ToDo: #1}}
\newcommand{\tday}{T_{\rm day}}
\newcommand{\tdaymax}{T_{\rm day,max}}
\newcommand{\tirr}{T_{\rm irr}}
\newcommand{\tirrroche}{T_{\rm irr,\,Roche}}
\newcommand{\deltar}{\Delta\mathcal{R}}

\author[0000-0003-0525-9647]{Zifan Lin (林梓帆)}
\affiliation{Department of Physics and McDonnell Center for the Space Sciences, Washington University, St. Louis, MO 63130, USA}
\email{lzifan@wustl.edu}

\author[0000-0002-6939-9211]{Tansu Daylan}
\affiliation{Department of Physics and McDonnell Center for the Space Sciences, Washington University, St. Louis, MO 63130, USA}
\email{tansu@wustl.edu}


\begin{abstract} 

Observing the dayside thermal emissions of rocky exoplanets provides essential insights into their compositions and the presence of atmospheres. 
Even though no conclusive evidence has been found for atmospheres on small rocky exoplanets orbiting M dwarfs, recent JWST observations identified puzzling thermal emission excesses. Some rocky exoplanets orbiting M dwarfs have dayside emission temperatures higher than the theoretical maximum temperatures, which are calculated assuming stellar irradiation as the sole energy source. Therefore, the observed thermal emission excesses imply that these planets may have internal heat sources.
In this work, we simulate three possible internal planetary processes that may generate excess heat in addition to stellar irradiation: residual heating from formation, tidal heating, and induction heating due to interactions with the stellar magnetic field.
We found that these mechanisms, even when combined, cannot explain the observed thermal emission excesses, nor can they account for a tentative positive trend in the brightness temperature scaling factor with irradiation temperature.
Our results imply that planetary internal processes are unlikely to generate remotely detectable heat, so the observed thermal excesses, if astrophysical, are likely caused by stellar contamination, surface processes, geometric effect, or other internal processes not considered in this study.
The ongoing JWST-HST Rocky Worlds Director's Discretionary Time Program and the upcoming Nancy Grace Roman Space Telescope will provide more insights into the thermal emission of rocky exoplanets.

\end{abstract}


\section{Introduction} \label{sect:intro}

Detecting atmospheres on small terrestrial exoplanets is among the foremost science goals of exoplanetary science. A prime approach to search for an atmosphere using emission spectroscopy is to look for \textit{thermal deficit}, the drop in the dayside brightness temperature ($\tday$) relative to the theoretically maximum dayside disk-integrated brightness temperature ($\tdaymax$) assuming zero albedo and no atmospheric redistribution of heat. Due to space weathering, bare rocks likely have darkened surfaces \citep{zieba_no_2023, lyu_super-earth_2024}. Therefore, if $\tday$ is significantly lower than $\tdaymax$, it is indicative of an atmosphere that redistributes heat from the permanent dayside to the nightside of a tidally locked planet. Such a thermal deficit was detected for planets orbiting K-type or hotter stars. 55 Cnc e, a large ($1.95\, R_\oplus$) and massive ($8.8\, M_\oplus$) planet orbiting a G8 star, has a low bulk density and an emission spectrum inconsistent with blackbody, both suggesting the presence of a volatile envelope \citep{hu_secondary_2024}. Thermal deficit was also recently detected on the ultra-hot rocky exoplanet TOI-561 b, which orbits a G9 star, implying the presence of a thick volatile atmosphere \citep{teske_thick_2025}. Spitzer observations revealed that rocky planets K2-141 b and TOI-431 b, both orbiting K-type stars, also have thermal deficits \citep{zieba_k2_2022, monaghan_low_2025}.

However, no thermal deficit has been detected on rocky exoplanets orbiting M dwarfs (``M-Earths''). Despite extensive searches since JWST's commissioning, conclusive evidence for an M-Earth atmosphere remains elusive \citep[e.g.,][]{kreidberg_first_2025}.
Recent JWST thermal phase curve results suggested that TRAPPIST-1 b is likely to be a bare rock \citep{greene_thermal_2023, gillon_first_2025, ducrot_combined_2025}. At the same time, the dayside brightness temperature ($\tday$) of TRAPPIST-1 c is still consistent with a tenuous atmosphere \citep{zieba_no_2023, gillon_first_2025}. The JWST Hot Rocks Survey (GO 3730, PI: Diamond-Lowe), which targeted nine highly irradiated rocky exoplanets, has not yet found a planet with an atmosphere \citep{august_hot_2025, valdes_hot_2025, fortune_hot_2025, allen_hot_2025}. The JWST-HST joint Rocky Worlds Director's Discretionary Time (DDT) Program will spend 500 hours searching for evidence for M-Earth atmospheres.  GJ 3929 b, one of the Rocky Worlds DDT targets, was observed by JWST during secondary eclipse \citep{xue_jwst_2025}. The authors reported a $\tday$ within the error of $\tdaymax$, implying that the planet is a dark bare rock. 

Intriguingly, some M-Earths have been found to exhibit \textit{thermal excess}: their measured $\tday$ values are marginally hotter than $\tdaymax$ predicted under the assumption that stellar irradiation is the only energy source. 
The observed thermal excess, if indeed astrophysical, implies that these rocky exoplanets may have internal heat sources capable of generating strong, remotely detectable surface heat flux. Among M-Earths observed to date, TOI-1468 b has the largest thermal excess \citep{valdes_hot_2025}, while LHS 1140 c \citep{fortune_hot_2025}, GJ 3929 b \citep{xue_jwst_2025}, and TOI-1685 b \citep{luque_dark_2025} also have $\tday$ marginally hotter than $\tdaymax$. 
Recently, \cite{coy_population-level_2025} identified a positive correlation between the dayside emission temperatures of M-Earths and their irradiation received from host stars, implying that the most irradiated M-Earths also have the strongest internal heating.

Several planetary interior mechanisms can potentially explain the observed thermal excess, namely (i) residual heat from formation that is retained by atmospheres, (ii) tidal heating, and (iii) induction heating. Below, we introduce each mechanism separately.

Residual heat from planet formation can be retained by a thick volatile envelope, which provides efficient thermal blanketing. Residual heat can maintain global magma oceans (MOs), which are expected to be common on planets that once had thick volatile envelopes \citep{vazan_contribution_2018, kite_atmosphere_2020}. As MOs cool down, they can emit surface heat fluxes as high as $\sim10^7$ W m$^{-2}$, and drop to $\sim10^0$ W m$^{-2}$ in $\gtrsim10^2$ Myr \citep{zhang_thermal_2022}. Using coupled planetary interior, thermal evolution, and atmospheric escape models, \cite{lin_most_2025} found that global MOs can be maintained on rocky exoplanets for $\sim1$ Gyr as photoevaporation gradually erodes their thick volatile envelopes. 
Therefore, if a rocky exoplanet is observed soon after total atmospheric loss, surface heat flux from an MO that is still cooling down can raise $\tday$ above $\tdaymax$, producing thermal excess.

Tidal dissipation as a heat source is common in the Solar System. Tidal heating is sensitive to orbital eccentricity and distance.
Close-in, eccentric exoplanets therefore experience significant tidal heating \citep[see][and references therein]{driscoll_tidal_2015}. Most observed M-Earths are very close to their host stars ($a\sim0.01$ AU), so if nonzero eccentricities can be maintained by mechanisms such as resonance with companions, thermal excess generated by tides is expected. Tidal response of a rocky planet with a liquid mantle layer may be amplified by a runaway tidal melting mechanism proposed by \cite{seligman_potential_2024}.
Elevated tidal response can lead to excessive dayside brightness temperature and violent volcanism. \cite{bello-arufe_evidence_2025} reported tentative evidence for a volcanic atmosphere on L 98-59 b, an M-Earth with $e=0.031^{+0.017}_{-0.016}$ \citep{cadieux_detailed_2025} in a multiplanet system. The high eccentricity of L 98-59 b is likely maintained by dynamical interactions with outer companions.

Induction heating is possible when a planet orbits a magnetically active star with an orbital axis tilted relative to the star's magnetic dipole axis, so that it experiences a changing external field. The changing field induces a current in the conductive internal layers of the planet, generating heat \citep{kislyakova_magma_2017, kislyakova_effective_2018, kislyakova_induction_2023}. Induction heating is significant for M-Earths, because fully convective, rapidly rotating M dwarfs are known to have strong (on the order of kG) dipolar fields \citep{shulyak_strong_2017}. In addition, magnetic flux carried by coronal mass ejections from active M dwarfs can lead to ohmic dissipation heating within a planet \citep{grayver_interior_2022}. Because M dwarfs have frequent flares \citep{gunther_stellar_2020, WhitsettDaylan2025}, this intermittent heating mechanism may have effects on the observed dayside emission temperatures of M-Earths. 

While previous literature has discussed possible causes of thermal excess in a subset of rocky exoplanets based on thermal emission observations \citep{coy_population-level_2025}, a systematic modeling effort to quantify the thermal excess potential of all rocky exoplanets is lacking. Here, thermal excess potential refers to a planet's ability to generate an excessive surface heat flux due to residual, tidal, or induction heating. In this work, we fill this gap by systematically modeling internal heating processes. We identify planets with high thermal excess potential, which are ideal targets for future thermal emission observations aiming at revealing internal physical mechanisms and evolutionary history of rocky exoplanets.

In this work, we will first analyze emission observations of rocky exoplanets to date to test the robustness of thermal emission excess measurements and to identify any trend in dayside emission temperatures. Then, we systematically model internal heating for all rocky exoplanets in the NASA Exoplanet Archive \citep{christiansen_nasa_2025} to identify the planets with the greatest thermal excess potentials. The rest of this paper is structured as follows. We introduce our modeling approaches in Section \ref{sect:methods}. The results are summarized in Section \ref{sect:results}. Section \ref{sect:discussion} discusses other mechanisms that may lead to thermal emission excess that we do not study in this work, and finally Section \ref{sect:conclusion} presents our findings.

\section{Methods} \label{sect:methods}

\subsection{Analyzing Past Emission Observations} \label{sect:methods_analyze_obs}

To test the robustness of the reported thermal emission excess and to identify any trend in $\mathcal{R}$, we collect rocky exoplanet emission observations in the literature and analyze them. Our data set includes 14 M-Earths: TRAPPIST-1 b, TRAPPIST-1 c, LHS 1140 c, LTT 1445 A b, GJ 3929 b, GJ 1132 b, TOI-1468 b, GJ 486 b, LHS 3844 b, LTT 3780 b, TOI-1685 b, GJ 367 b, GJ 1252 b, and LHS 1478 b. For completeness, our dataset further includes four rocky exoplanets orbiting K- or G-type stars: 55 Cnc e, K2-141 b, TOI-431 b, and TOI-561 b. Table \ref{tab:obs_planets_data} summarizes the $\tday$ and the temperature scaling factor $\mathcal{R}$ of rocky exoplanets observed to date. The factor $\mathcal{R}$ is defined by \cite{xue_jwst_2024} and \cite{weiner_mansfield_no_2024} to quantify the ratio of measured dayside temperature $\tday$ to the theoretical maximum disk-integrated dayside temperature $\tdaymax$
\begin{equation}
    \mathcal{R} \equiv \frac{\tday}{\tdaymax}.
\end{equation}
By definition, $\mathcal{R}\leq1$ if the stellar irradiation is the only heat source. $\mathcal{R}<1$ implies a reflective surface with albedo $A_B > 0$, the presence of an atmosphere that transports heat (i.e., heat redistribution efficiency $\varepsilon>0$) from the permanent dayside to the nightside of a tidally locked planet, or a combination of both effects. $\mathcal{R}>1$, which this work aims to investigate, implies planetary internal heat sources.

Note that some planets above, such as 55 Cnc e, have low bulk densities that are consistent with some volatile envelopes, and are therefore not purely rocky. We define the criteria of being ``rocky'' in Section \ref{sect:methods_rocky_selection}.

\begin{deluxetable*}{lllccl}
\tabletypesize{\scriptsize}
\tablewidth{0pt}
\tablecaption{Rocky exoplanet dayside emission observed to date \label{tab:obs_planets_data}}
\tablehead{
\colhead{Planet} & \colhead{Instrument} & \colhead{Type\tablenotemark{a}} & \colhead{$\tday$ (K)} & \colhead{$\mathcal{R}$} & \colhead{Reference}
}
\startdata
\textbf{M-Earths} \\ \hline
GJ 1132 b          & MIRI/LRS      & E    & $709\pm31$       & $0.95\pm0.04$      & \cite{xue_jwst_2024}         \\ \hline
GJ 1252 b          & Spitzer       & E    & $1410^{+91}_{-125}$  & $1.01\pm0.09$      & \cite{crossfield_gj_2022}  \\ \hline
GJ 367 b           & MIRI/LRS      & PC   & $1728\pm90$      & $0.97\pm0.10$      & \cite{zhang_gj_2024}      \\ \hline
GJ 3929 b\tablenotemark{b} & MIRI/F1500W   & E    & $745\pm76$       & $1.01\pm0.10$      & \cite{xue_jwst_2025}         \\ 
GJ 3929 b\tablenotemark{c}   & MIRI/F1500W   & E    & $782\pm79$       & $1.07\pm0.10$      & \cite{xue_jwst_2025}         \\ \hline
GJ 486 b           & MIRI/LRS      & E    & $865\pm14$       & $0.97\pm0.01$      & \cite{weiner_mansfield_no_2024}   \\ \hline
LHS 1140 c         & MIRI/F1500W   & E    & $561\pm44$       & $1.04\pm0.08$      & \cite{fortune_hot_2025}     \\ \hline
LHS 1478 b\tablenotemark{d}         & MIRI/F1500W   & E    & $491\pm102$      & $0.64\pm0.13$      & \cite{august_hot_2025}      \\ \hline
LHS 3844 b         & Spitzer       & PC   & $1040\pm40$      & $1.01\pm0.05$      & \cite{kreidberg_absence_2019}   \\ \hline
LTT 1445 A b       & MIRI/LRS      & E    & $525\pm15$       & $0.952\pm0.057$    & \cite{wachiraphan_thermal_2025} \\ \hline
LTT 3780 b         & MIRI/F1500W   & E    & $1143^{+104}_{-99}$  & $0.98\pm0.09$      & \cite{allen_hot_2025}       \\ \hline
TOI-1468 b         & MIRI/F1500W   & E    & $1024^{+78}_{-74}$   & $1.17\pm0.10$      & \cite{valdes_hot_2025}      \\ \hline
TOI-1685 b (NRS1)    & NIRSpec/G395H & PC   & $1520\pm140$     & $1.10\pm0.10$      & \cite{luque_dark_2025}       \\
TOI-1685 b (NRS2)    & NIRSpec/G395H & PC   & $1360^{+100}_{-90}$  & $0.99\pm0.07$      & \cite{luque_dark_2025}       \\ \hline
TRAPPIST-1 b       & MIRI/F1280W   & E    & $424\pm28$       & $0.83\pm0.06$      & \cite{ducrot_combined_2025}      \\
                   & MIRI/F1500W   & E    & $478\pm27$       & $0.94\pm0.05$      & \cite{ducrot_combined_2025}      \\
                   & MIRI/F1500W   & PC   & $490\pm17$       & $0.96\pm0.04$      & \cite{gillon_first_2025}      \\ \hline
TRAPPIST-1 c       & MIRI/F1500W   & E    & $380\pm31$       & $0.88\pm0.07$      & \cite{zieba_no_2023}       \\
                   & MIRI/F1500W   & PC   & $369\pm23$       & $0.85\pm0.05$      & \cite{gillon_first_2025} \\ \hline
\textbf{K \& G hosts} \\ \hline
55 Cnc e           & MIRI/LRS      & E    & $1796\pm88$      & $0.71\pm0.04$      & \cite{hu_secondary_2024}          \\ \hline
K2-141 b           & K2; Spitzer   & PC   & $2049^{+362}_{-359}$ & $0.75\pm0.13$ & \cite{zieba_k2_2022}       \\ \hline
TOI-431 b          & Spitzer       & E    & $1520^{+360}_{-390}$ & $0.63\pm0.16$      & \cite{monaghan_low_2025}    \\ \hline
TOI-561 b (\texttt{Eureka!})  & NIRSpec/BOTS  & E    & $1740^{+70}_{-80}$   & $0.59\pm0.03$      & \cite{teske_thick_2025}       \\
TOI-561 b (\texttt{JEDI 1})   & NIRSpec/BOTS  & E    & $1830\pm70$      & $0.62\pm0.03$      & \cite{teske_thick_2025}       \\
TOI-561 b (\texttt{JEDI 2})   & NIRSpec/BOTS  & E    & $2150\pm80$      & $0.73\pm0.03$      & \cite{teske_thick_2025}       \\
\enddata
    \tablecomments{TOI-1685 b has two sets of reported values from the NRS1 and NRS2 wavelength ranges of JWST/NIRSpec. TOI-561 b has three sets of values from different reduction pipelines. \tablenotetext{a}{``Type'' refers to whether the observation was secondary eclipse (E) or phase curve (PC).} \tablenotetext{b}{Data derived from observed stellar flux.} \tablenotetext{c}{Data derived from SPHINX stellar model.} \tablenotetext{d}{Excluded from analysis (see Section \ref{sect:methods_analyze_obs}).}}
\end{deluxetable*}

We exclude LHS 1478 b from our analysis. The calculated $\mathcal{R}=0.64\pm0.13$ for LHS 1478 b is significantly lower than all other M-Earths, making it an outlier. \cite{august_hot_2025} reported that their two visits did not yield consistent results, where the second visit was strongly affected by correlated noise. \cite{connors_uniform_2025} also reported that they were unable to detect an eclipse from the second visit, confirming the earlier claim by \cite{august_hot_2025}.

Emission observations typically report the measured dayside brightness temperature, $\tday$. If the entire orbital phase is covered, the nightside temperature, $T_{\rm night}$, is also reported \citep[e.g.,][]{Daylan+2021b}. Theoretically, the maximum temperature on a planet's surface that can be reached, assuming stellar irradiation is the sole energy source, is the irradiation temperature, i.e., the equilibrium temperature at the substellar point
\begin{equation} \label{eq:tirr_def}
    \tirr = \frac{T_{\rm eff}}{\sqrt{a/R_*}} ,
\end{equation}
where $T_{\rm eff}$ is the host star's effective temperature, $a$ the orbital semi-major axis, and $R_*$ the stellar radius. According to \cite{cowan_statistics_2011}, the disk-integrated dayside brightness temperature is related to $\tirr$ by
\begin{equation} \label{eq:tday_def}
    \tday = \tirr \cdot (1-A_B)^{1/4} \cdot \left(\frac{2}{3} - \frac{5}{12}\varepsilon\right)^{1/4},
\end{equation}
where $A_B$ is the Bond albedo and $0\leq\varepsilon\leq1$ is the heat redistribution efficiency. In the limit of $A_B \rightarrow 0$ and $\varepsilon\rightarrow0$, i.e., no heat gets reflected or redistributed away from the dayside, the theoretically maximum dayside temperature, $\tdaymax$, is reached, where
\begin{equation}
    \tdaymax = \left(\frac{2}{3}\right)^{1/4} \tirr.
\end{equation}

To identify any trend in $\mathcal{R}$ as a function of $\tirr$, we perform two types of fits: linear fit and broken power law (BPL) fit. The linear model takes the standard form of linear regression
\begin{equation}
    \mathcal{R} = A + B \cdot \tirr,
\end{equation}
where the constants $A$ and $B$ are fitted using \texttt{scipy.odr}, an orthogonal distance regression (ODR) routine that can take uncertainties in both the dependent and independent variables into account. For the BPL fit, we assume a functional form of 
\begin{equation}
\mathcal{R} =
\begin{cases}
C \left( \dfrac{\tirr}{T_b} \right)^{-\alpha_1}, & \text{if } x \leq x_b \\
C \left( \dfrac{\tirr}{T_b} \right)^{-\alpha_2}, & \text{if } x > x_b
\end{cases}
\end{equation}
where the constants $C$, $\alpha_1$, $\alpha_2$ and the break point temperature $T_b$ are also fitted with \texttt{scipy.odr}.

We adopt a step function as the null hypothesis, where $\mathcal{R} = 0.983$ for M-Earths ($\tirr \leq 2500$ K), and $\mathcal{R} = 0.671$ for other rocky exoplanets ($\tirr > 2500$ K). We choose $\sim2500$ K as the switch point because it is the maximum $\tirr$ an M-Earth can have before falling within the Roche limit (see Section \ref{sect:methods_id_hottest}). After performing linear and BPL fits, we quantify the goodness of the fits relative to the null hypothesis by performing reduced $\chi^2$ analysis. 

\subsection{Target Selection: All Rocky Exoplanets} \label{sect:methods_rocky_selection}

To identify the rocky exoplanets with the highest thermal excess potential, we searched the NASA Exoplanet Archive \citep {christiansen_nasa_2025} for all confirmed exoplanets orbiting single stars with known masses and radii as of October 22, 2025. Because there may be multiple entries for the same exoplanet, and the NASA Exoplanet Archive composite data may not be self-consistent if coming from different sources, we further consult the PlanetS catalog \citep{otegi_revisited_2020, parc_super-earths_2024}. The PlanetS catalog contains planets with well-characterized masses and radii ($\delta M_p/M_p\leq25\%$, $\delta R_p/R_p\leq8\%$) and chooses the most precise and recent reference among all reported values. Hence, if an exoplanet is in the PlanetS catalog, we adopt the PlanetS planet and system parameters for that exoplanet. We remove planets that are flagged as controversial or circumbinary in the NASA Exoplanet Archive. We also remove planets without valid $M_p$, $\delta M_p$, $R_p$, or $\delta R_p$ values. If there are multiple entries for a single planet, we choose the entry with the smallest $\delta M_p/M_p$ and $\delta R_p/R_p$. In total, our dataset contains 1289 planets and is available on Zenodo \citep{LinDaylan2026}. We provide documentation and functionality to reproduce our target selection and build upon our decomposition of thermal excess via the Python package \texttt{calor}, available on GitHub\footnote{https://github.com/astromusers/calor}.

\begin{figure}[t]
    \centering
    \includegraphics[width=\linewidth]{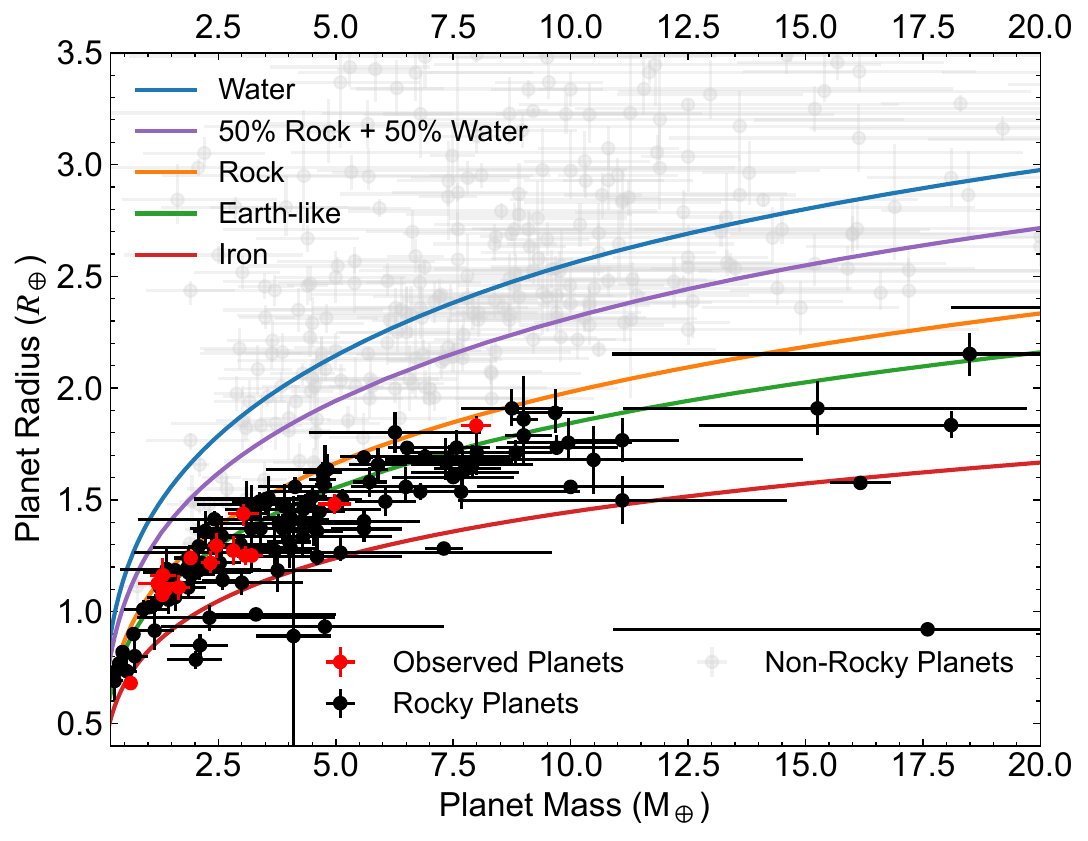}
    \caption{Mass-radius plot for exoplanets studied in this work. Rocky ($p_{\rm rocky}\geq32\%$) planets are shown in black, while non-rocky planets are shown in gray. Observed rocky exoplanets (Table \ref{tab:obs_planets_data}) are highlighted in red. Constant composition curves calculated using \cite{zeng_massradius_2016} rock (MgSiO$_3$) and iron equations of state, as well as the AQUA water equation of state \citep{haldemann_aqua_2020}, are plotted for comparison.}
    \label{fig:rocky_planets_MR}
\end{figure}

The interior compositions of exoplanets suffer from an intrinsic degeneracy \citep[e.g.,][]{valencia_detailed_2007, zeng_computational_2008, rogers_framework_2010}, so it is difficult to confidently conclude whether an exoplanet is rocky or not. Here, we adopt a probabilistic definition of ``rocky'' similar to how \cite{cambioni_can_2025} defined high-density exoplanets. For each exoplanet, we sample $10^5$ mass-radius (M-R) pairs within the uncertainty ellipse, and compute the fraction of those points that fall below the pure silicate M-R curve. This fraction is defined as $p_{\rm rocky}$. The pure silicate M-R curve is computed using the planetary interior model \texttt{CORGI}\footnote{https://github.com/Zifan-Lin/CORGI} \citep{lin_interior_2025}, adopting the \cite{zeng_massradius_2016} silicate mantle equation of state. A planet is declared ``rocky'' if more than 32\% of M-R pairs are below the pure silicate curve, i.e., the planet is not $>1\sigma$ less dense than a purely rocky composition. In total, there are 137 rocky exoplanets in our dataset (Figure \ref{fig:rocky_planets_MR}), among which 18 have been observed for their dayside thermal emission (Table \ref{tab:obs_planets_data}).

\subsection{Identifying the Hottest Planets} \label{sect:methods_id_hottest}

\begin{figure}[t]
    \centering
    \includegraphics[width=\linewidth]{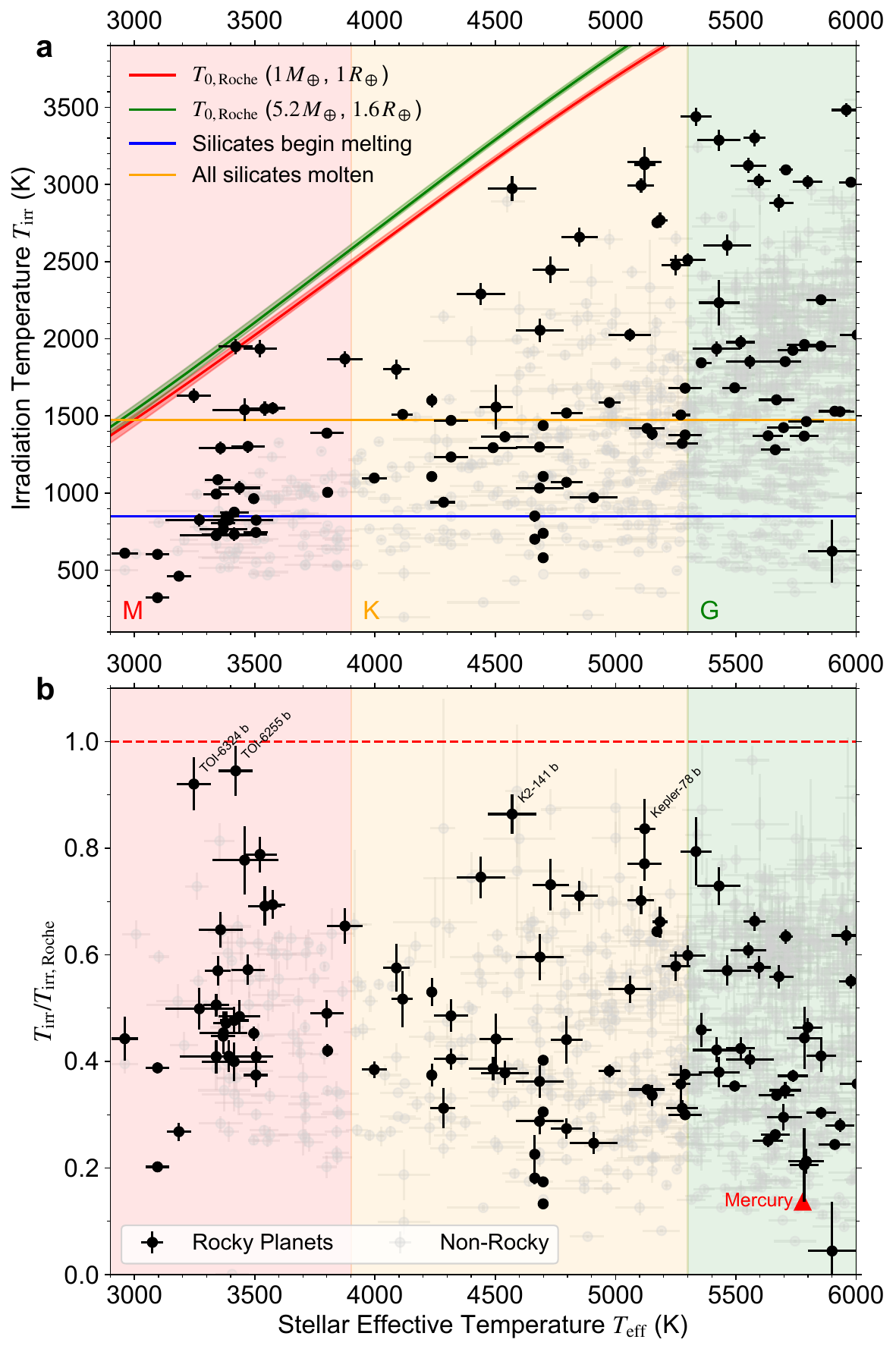}
    \caption{(a) Irradiation temperature $\tirr$ and (b) ratio of $\tirr$ to the maximum irradiation limited by the Roche distance, $\tirr/\tirrroche$, as functions of stellar effective temperature. In (a), two different $\tirrroche$ limits assuming Earth-like (red) and super-Earth (green) compositions are shown. Temperatures at which silicates begin to melt (850 K) and are fully molten (1473 K) are shown for comparison \citep{lutgens_essentials_2015}. In (b), Mercury is shown as a red triangle for reference, and exoplanets closest to Roche limits are annotated. Planets near the Roche limit are ideal targets for the search for thermal emission excesses.}
    \label{fig:tirr_vs_tirrRoche}
\end{figure}

The surface temperature of a planet cannot be arbitrarily hot, because it cannot be arbitrarily close to its host star. The Roche limit distance ($a_{\rm Roche}$) for a planet composed of incompressible fluid with negligible bulk tensile strength is \citep{rappaport_roche_2013}
\begin{equation}
    a_{\rm Roche} \simeq 2.44 R_p \left( \frac{M_*}{M_p} \right)^{1/3},
\end{equation}
where we rewrite it in terms of stellar and planetary masses instead of densities for convenience. Substituting $a_{\rm Roche}$ into Eq. (\ref{eq:tirr_def}) gives the irradiation temperature at the Roche limit, $T_{\rm irr,\, Roche}$, which is the maximum $\tirr$ a planet can have before tidal disintegration. If the positive $\mathcal{R}(\tirr)$ trend identified by \cite{coy_population-level_2025} is real, planets with $\tirr$ close to $\tirrroche$ are ideal targets for the search for thermal emission excess.

Figure \ref{fig:tirr_vs_tirrRoche} shows rocky exoplanets $\tirr$ compared to $\tirrroche$, where the planets are color coded by composition (see Section \ref{sect:methods_rocky_selection}). Figure \ref{fig:tirr_vs_tirrRoche}a shows $\tirr$ compared to two different $\tirrroche$ curves assuming Earth-like ($1\,M_\oplus$, $1\,R_\oplus$, red curve) and super-Earth ($5.2\,M_\oplus$, $1.6\,R_\oplus$, green curve) compositions. Figure \ref{fig:tirr_vs_tirrRoche}b shows the $\tirr/\tirrroche$ ratio, which is calculated based on the individual planetary and stellar parameters of each system. 

A stellar $M_*$-$R_*$-$T_{\rm eff}$ relation is needed to generate the above $\tirrroche$ curves. $M_*$-$R_*$ relations of cool ($\lesssim6000$ K) stars in our dataset generally follow a linear relation, which we use \texttt{scipy.odr} to find the best fit to be $(M_*/M_\odot) = 1.110 (R_*/R_\odot) - 0.046$.
Following \cite{mann_how_2015}, we fit an empirical third-order polynomial relation $(R_*/R_\odot) = a + bT_{\rm eff} +cT_{\rm eff}^2+dT_{\rm eff}^3$ to host stars in our database. The best-fit parameters are $a=-12.667$, $b=8.923\times10^{-3}$, $c=-2.022\times10^{-6}$, and $d=1.551\times10^{-10}$.

Several rocky exoplanets that orbit their host stars near the Roche distance are annotated in Figure \ref{fig:tirr_vs_tirrRoche}b. In particular, tidal orbital decay may cause TOI-6255 b to fall within the Roche limit in $\sim400$ Myr \citep{dai_earth-sized_2024}. Similarly, TOI-6324 b is predicted to reach the Roche limit in $\sim550$ Myr \citep{lee_toi-6324_2025}. They are therefore among the most promising M-Earths for validating the trend in thermal emission excess. A JWST/MIRI phase curve observation is planned for TOI-6255 b (GO 8864, PI: Lisa Dang).

\subsection{Modeling Residual Heating}

We model the thermal emission excess of rocky exoplanets due to primordial heat retained by a volatile envelope in two steps. First, we calculate the timescale of the total loss of a hydrogen and helium (H/He) atmosphere due to photoevaporation ($\tau_{\rm loss}$). Then, we simulate the surface MO cooling and the resulting surface heat flux, given that the surface is no longer insulated by an atmosphere.

To calculate the photoevaporation atmospheric loss timescale $\tau_{\rm loss}$, we adopt the analytical model by \cite{owen_evaporation_2017}, which was previously applied to simulate the envelope loss of rocky exoplanets that were once shrouded by massive volatile envelopes \citep{lin_most_2025}
\begin{equation} \label{eq:tau_loss}
\begin{split}
    \tau_{\rm loss} &= 210 \text{ Myr } \left(\frac{\eta}{0.1}\right)^{-1} \left(\frac{L_{\rm HE}}{10^{-3.5} L_\odot}\right)^{-1} \\
    &\times \left(\frac{P}{10 \text{ days}}\right)^{1.41} \left(\frac{M_*}{M_\odot}\right)^{0.52} \left(\frac{f}{1.2}\right)^{-3} \\
    &\times \left(\frac{\tau_{\rm KH}}{100 \text{ Myr}}\right)^{0.37} \left(\frac{\rho_c}{5.5 \text{ g cm}^{-3}}\right)^{0.18} \left(\frac{M_c}{5M_\oplus}\right)^{1.42} \\
    &\times \begin{cases}
        \left(\frac{\Delta R}{R_c}\right)^{1.57} & \text{ if } \Delta R / R_c < 1 \\
        \left(\frac{\Delta R}{R_c}\right)^{-1.69} & \text{ if } \Delta R / R_c \geq 1,
    \end{cases}
\end{split}
\end{equation}
where $\eta$ is the dimensionless efficiency factor. Typically, $\eta$ is assumed to be a constant ranging from $\sim0.1$--0.3 \citep{lopez_how_2012, owen_evaporation_2017}, which is convenient but may not apply to all planets. Hence, we use the analytical approach proposed by \cite{caldiroli_irradiation-driven_2022} to estimate $\eta$. $L_{\rm HE}$ is the high-energy luminosity of the host star, assuming that $L_{\rm HE} = 10^{-3.5}L_{\rm bol}$, where $L_{\rm bol}$ is the bolometric luminosity of the star. $P$ is the planetary orbital period. $M_*$ is stellar mass. $f$ is a radius scaling factor, such that $R_p = f R_{\rm RCB}$, where $R_{\rm RCB}$ is the radius of the radiative-convective boundary. For simplicity, we assume that $f=1$. $\tau_{\rm KH}$ is the Kelvin-Helmholtz timescale, typically assumed to be the larger of 100 Myr or the planet's age \citep{owen_evaporation_2017}. $\rho_c$ is the density of a $1\, M_\oplus$ core -- here, ``core'' means the iron-silicate part of the planet surrounded by the volatile envelope -- which equals 5.5 g cm$^{-3}$ for an Earth-like composition. $M_c$ is the core mass. $\Delta R/R_c$ is the ratio of envelope thickness to core radius.

The atmospheric loss timescale $\tau_{\rm loss}$ depends on the initial atmospheric mass fraction $X$, and peaks at $X\sim0.01$ for a $5\, M_\oplus$ super-Earth with an Earth-like interior composition orbiting a Sun-like star \citep{owen_evaporation_2017}. For consistency, we assume that all planets have an initial envelope mass equaling 1\% of planetary mass (i.e., $X=0.01$).

Using a combined planetary interior and thermal evolution model, \cite{lin_most_2025} showed that both hydrogen/helium- and water-dominated envelopes are highly efficient in keeping the core in a hot and molten state. Hence, soon after total atmospheric loss, a rocky exoplanet should have extensive surface MO given a suitable equilibrium temperature, $T_{\rm eq}$.

To model the time evolution of surface heat flux after total atmospheric loss, we use a 1D rocky exoplanet thermal evolution code \texttt{CMAPPER}.\footnote{\href{https://github.com/zhangjis/CMAPPER_rock}{https://github.com/zhangjis/CMAPPER\_rock}} Detailed descriptions of an early version of the model can be found in \cite{zhang_thermal_2022}. The surface heat flux is assumed to be
\begin{equation}
    F_{\rm surf} = \sigma(T_{\rm surf}^4 - T_{\rm eq}^4),
\end{equation}
where the surface temperature $T_{\rm surf}$ is calculated from processes including planetary internal heat production, heat transport across the core-mantle boundary, and mantle heat transport. For this study, we assume the default mantle equation of state and simulate the time evolution of $F_{\rm surf}$ to investigate if it can explain the observed thermal emission excess. Because \texttt{CMAPPER} is limited to rocky planets with $0.5\, M_\oplus\leq M_p \leq8\, M_\oplus$, core mass fraction (CMF) between 0.1 and 0.7, and 255 K $\leq T_{\rm eq} \leq2700$ K, we only model the residual thermal evolution of a subset of 68 planets.

\subsection{Modeling Tidal Heating}

Tidal heating rates of exoplanets orbiting their host stars on eccentric orbits can be approximated by the ``fixed-$\mathcal{Q}$'' model \citep{driscoll_tidal_2015}. The total tidal power, $Q_{\rm tidal}$, is calculated as
\begin{equation}
    Q_{\rm tidal} = -\frac{21}{2} {\rm Im}(k_2) G^{3/2} M_*^{5/2} R_p^5 \frac{e^2}{a^{15/2}}, 
\end{equation}
where ${\rm Im}(k_2)$ is the imaginary part of the second-order Love number $k_2$. ${\rm Im}(k_2)$ is commonly approximated as ${\rm Im}(k_2) \approx{\rm Re}(k_2)/\mathcal{Q}$, where ${\rm Re}(k_2)$ is the real part of $k_2$ and $\mathcal{Q}$ the tidal quality factor. Present-day Earth has an empirical $-{\rm Im}(k_2) = 0.003$, where ${\rm Re}(k_2)=0.3$ and $\mathcal{Q}\approx100$ \citep{ray_fortnightly_2012}. The tidal response of Io is five times stronger than Earth, with $-{\rm Im}(k_2) = 0.015\pm0.003$, ${\rm Re}(k_2)\approx0.04$, and $\mathcal{Q}\approx3$ under typical mantle rigidity and viscosity assumptions \citep{lainey_strong_2009}. We adopt Earth-like and Io-like $-{\rm Im}(k_2)$ values to account for the different tidal responses of rocky exoplanets with different compositions and interior melting states. The surface tidal heat flux is $F_{\rm tidal} = Q_{\rm tidal}/4\pi R_p^2$.

Tides raised on a planet by its host star can circularize its orbit. To assess if a planet can maintain a nonzero eccentricity that is required for tidal heating, given its age, we calculate the tidal circularization timescale, $\tau_{\rm circ}$, which is defined as \citep{jackson_tidal_2008}
\begin{equation}
    \tau_{\rm circ} = \left[\frac{63}{4}(GM_*^3)^{1/2}\frac{R_p^5}{\mathcal{Q}_p M_p}\right]^{-1} a^{13/2}.
\end{equation}

Because determining the ages of M dwarfs is highly challenging \citep{engle_living_2023}, many M-Earths in our database lack reliable age estimates. In this case, $\tau_{\rm circ}$ should be compared to typical planet system ages (on the order of several Gyr). Even if $\tau_{\rm circ}$ is smaller than planet age, it does not imply that the planet cannot maintain a nonzero eccentricity at the present day, because dynamical interactions with undetected companions are still possible.

\subsection{Modeling Induction Heating}

We use the analytical model outlined in \cite{kislyakova_magma_2017} to calculate induction heating. The energy released within the planet by induction heating is
\begin{equation}
    Q_{\rm induction} = \frac{1}{2\sigma} \int|j_\phi|^2 dV,
\end{equation}
where $\sigma$ is electrical conductivity and $j_\phi$ is the induced current. In the full analytical model, the planet is divided into many layers, each with a different conductivity, and $Q$ and $j$ are computed for each layer. Here, we adopt several simplifications because we only aim to derive the approximate induction heat flux for a large number of exoplanets, but do not attempt to model induction heating in the same amount of detail as previous works \citep{kislyakova_magma_2017, kislyakova_effective_2018}. The first simplification is that we assume the entire planet has uniform conductivity $\sigma = 5\times10^{10}$ (in CGS), a reasonable value for molten rocks, following \cite{kislyakova_effective_2018}. The second simplification is that we assume all planets have orbital inclinations of 90\textdegree, which implies they experience the largest possible external field variations as they orbit their host stars, thereby placing an upper limit on $Q_{\rm induction}$. The third simplification is that we assume all host stars have strong dipolar fields similar to that of WX UMa, with a strength of 7.3 kG \citep{shulyak_strong_2017}. Again, this assumption places an upper limit on $Q_{\rm induction}$ because such strong fields are only expected for young, rapidly rotating M dwarfs. Stars with multipolar fields are expected to have field strength below 4 kG \citep{shulyak_strong_2017}. These simplifications are not only convenient but also necessary. A reliable conductivity profile requires detailed planetary interior modeling that is beyond the scope of this study. The orbital inclinations of most exoplanets are unknown, with only a small subset having inclination measured using the Rossiter-McLaughlin effect \citep[e.g.,][]{frazier_neid_2023}. Magnetic field strength measurements for exoplanet host stars are also scarce.

To calculate the magnetic field at the planet's orbit, we follow the prescription of \cite{kislyakova_effective_2018}. We assume that the stellar magnetic field decreases as $r^{-3}$ from the stellar surface to the ``source surface'' at $R_{\rm ss}=2.5R_*$. Beyond the source surface, the field strength decreases at a slower rate of $r^{-2}$.

\subsection{Quantifying Thermal Excess Potential}

We define the thermal emission excess produced by a certain mechanism, $\Delta \mathcal{R}_i$, as \citep[following][]{coy_population-level_2025}
\begin{equation} \label{eq:deltaR_def}
    \deltar_i = \left(\frac{F_i + F_{\rm insol}}{F_{\rm insol}}\right)^{1/4}-1,
\end{equation}
where $F_{\rm insol}$ is the disk-averaged stellar insolation, and the subscript $i$ denotes a certain planetary internal heating mechanism. The total thermal excess potential is therefore
\begin{equation} \label{eq:total_r}
    \deltar_{\rm total} = \sum_i \deltar_i.
\end{equation}
In this work, we consider $\deltar_{\rm residual} + \deltar_{\rm tidal} + \deltar_{\rm induction}$, but the expression can be easily expanded in the future if more heating mechanisms are identified. Note that in \cite{coy_population-level_2025}, $\deltar_i$ was defined without the $-1$ term, which we introduced to emphasize that it is a delta value that quantifies an increase in thermal emission and to allow superposition of multiple heating mechanisms.

\section{Results} \label{sect:results}

\begin{figure*}[t]
    \centering
    \includegraphics[width=0.9\textwidth]{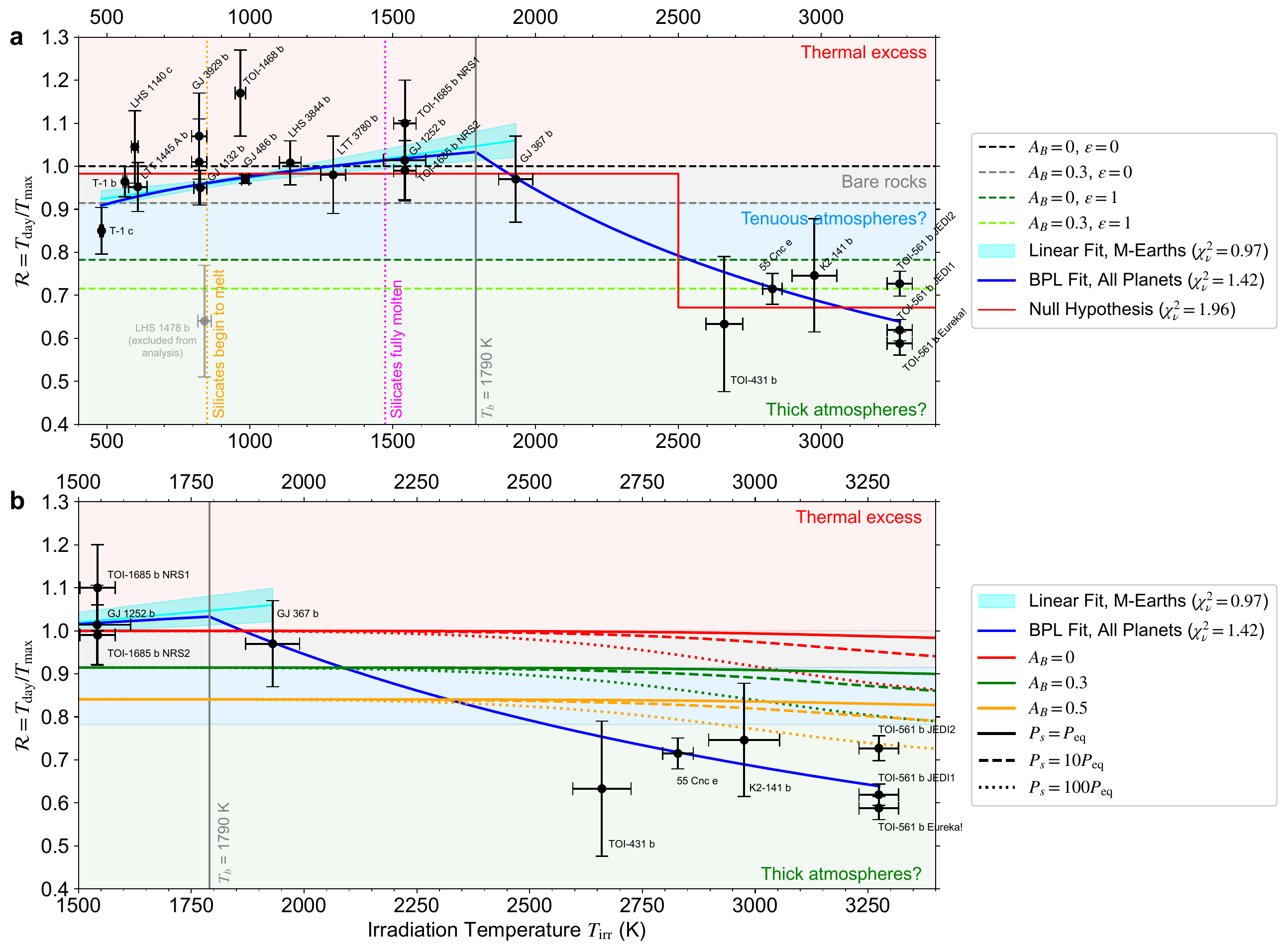}
    \caption{Dayside temperature scaling factor, $\mathcal{R}=\tday/\tdaymax$, as a function of irradiation temperature, $\tirr$, for (a) all observed rocky exoplanets and (b) planets with $\tirr>1500$ K, which may have rock vapor atmospheres. Observed exoplanets (Table \ref{tab:obs_planets_data}) are shown as black error bars, except for LHS 1478 b (gray error bar), which is an outlier and is excluded from our analysis (see Section \ref{sect:methods_analyze_obs}). Two fits are shown: (cyan) linear fit for M-Earths shows a positive trend, while (blue) BPL fit shows a positive trend for M-Earths with $\tirr<T_b=1790$ K and a negative trend for five extremely hot planets with $\tirr>T_b$. Both fits are better than the null hypothesis (red line in panel a), suggesting an underlying physical mechanism that produces both thermal emission excess and deficit. In (a), the melting temperatures of silicates (orange and magenta vertical lines) are annotated. In (b), $\mathcal{R}$ of exoplanets with rock vapor atmospheres derived from a scaling relation \citep{koll_scaling_2022} are shown in red, green, and orange for surface Bond albedo of 0, 0.3, and 0.5, respectively. Thick atmospheres with $P_s$ much greater than the equilibrium pressure $P_{\rm eq}$ are required to explain the observed thermal deficit.}
    \label{fig:tirr_vs_r_fits}
\end{figure*}

Here, we present the results on thermal emission models for rocky exoplanets. Section \ref{sect:res1_positive_trend} and \ref{sect:results_negative_trend} discuss the statistical trend found in the observed temperature scaling factor $\mathcal{R}$ as a function of $\tirr$. Section \ref{sect:res_residual_heating}--\ref{sect:res_induction_heating} presents the results for three possible internal heating mechanisms that may explain the observed thermal emission excess. Finally, in Section \ref{sect:res_combined}, we combine the results for various mechanisms and conclude that known internal processes are unlikely to produce the observed thermal emission excess.


\subsection{A Tentative Positive $\mathcal{R}(\tirr)$ Trend Implies Hotter M-Earths Produce More Thermal Excess} \label{sect:res1_positive_trend}

We find that a positive $\mathcal{R}(\tirr)$ trend is favored over the null hypothesis for M-Earths (reduced $\chi_\nu^2=0.97$, compared to the $\chi_\nu^2=1.96$ of the null hypothesis), according to the linear fit (Figure \ref{fig:tirr_vs_r_fits}). This suggests that, if the thermal emission excesses are indeed astrophysical, hotter rocky planets are more likely to produce higher excesses. The BPL fit also finds a positive trend for M-Earths up to a break point temperature $T_b=1790$ K, after which the trend becomes negative, implying increasing reflectivity and/or atmospheric pressure for hotter rocky exoplanets (see Section \ref{sect:results_negative_trend}). While not as good a fit as the linear fit, the BPL fit is also marginally favored over the null hypothesis (reduced $\chi_\nu^2=1.42$, compared to $\chi_\nu^2=1.96$). The constants for the linear fit are: $A=0.879\pm0.036$, $B=(9.372\pm3.775)\times10^{-5}$, which are broadly consistent with the previous linear fit by \cite{coy_population-level_2025}, who found $A=0.8431\pm0.0132$ and $B=(13\pm1.3)\times10^{-5}$. Note, however, that our M-Earth dataset includes four more planets (13 in total, compared to 9 M-Earths) than the previous fit.
The constants for the BPL fit are: $C=1.033 \pm 0.032$, $T_b=1790\pm256$ K, $\alpha_1=-0.097\pm0.043$, $\alpha_2=0.796\pm0.219$. The null hypothesis we adopt is $\mathcal{R}=0.983$ if $\tirr\leq2500$ K (i.e., M-Earths are all dark bare rocks) and $\mathcal{R}=0.671$ if $\tirr>2500$.
Figure \ref{fig:tirr_vs_r_fits} shows the $\mathcal{R}$ and $\tirr$ of all measured rocky exoplanets, as well as the linear fit, BPL fit, and null hypothesis. 

To test the robustness of the trend, we perform leave-one-out tests by performing fits while removing one planet from the sample at a time in Appendix \ref{sect:app_robustness_of_trend}.

\subsection{A Negative $\mathcal{R}(\tirr)$ Trend that Requires Thick Atmospheres and High Albedo to Explain} \label{sect:results_negative_trend}

In addition to the positive trend for M-Earths previously identified in the literature \citep{coy_population-level_2025}, we further find a negative trend for extremely hot rocky exoplanets with $\tirr > T_b = 1790$ K. This trend is driven by four planets orbiting K- or G-type stars, namely TOI-431 b \citep{monaghan_low_2025}, 55 Cnc e \citep{hu_secondary_2024}, K2-141 b \citep{zieba_k2_2022}, and TOI-561 b \citep{teske_thick_2025}. Given their low measured $\mathcal{R}$, these planets may have rock vapor atmospheres that are in equilibrium with surface MOs \citep{schaefer_chemistry_2009, miguel_compositions_2011} or outgassed atmospheres generated by gases released from the MOs.

To explore whether rock vapor atmospheres can drive the observed negative trend, we adopt the pressure-temperature relation from \cite{ito_theoretical_2015}, which displays a monotonic increase in the total pressure of an atmosphere in equilibrium with the MO ($P_{\rm eq}$) as $\tirr$ increases. Because \cite{ito_theoretical_2015} showed a limited temperature range ($1500\leq T\leq3000$ K), we fit the following empirical $P_{\rm eq}$-$\tirr$ relation for extrapolation
\begin{equation}
\begin{split}
    P_{\rm eq} = &-4.96\times10^{-13} \tirr^4 + 5.41\times10^{-9}\tirr^3 - \\
    &2.27\times10^{-5}\tirr^2 + 4.62\times10^{-2}\tirr - 4.06,
\end{split}
\end{equation}
where a fourth-order polynomial fit is chosen so that it most accurately reproduces the behavior of the $P_{\rm eq}$-$\tirr$ relation assuming a bulk silicate Earth magma composition. To account for magma outgassing that can lead to surface pressure $P_s>P_{\rm eq}$, we further consider two cases where the total atmospheric pressure is $10\times$ (dashed line) or $100\times$ (dotted line) $P_{\rm eq}$ (Figure \ref{fig:tirr_vs_r_fits}).

\cite{koll_scaling_2022} presented an analytical scaling relation that links a planet's heat redistribution to its equilibrium temperature, surface pressure, and broadband longwave optical thickness ($\tau_{\rm LW}$). This relation, validated against general circulation models, is
\begin{equation}
    \varepsilon = \frac{\tau_{\rm LW}^{1/3}\left(\frac{P_s}{\rm 1\,bar}\right)^{2/3}\left(\frac{T_{\rm eq}}{\rm 600\,K}\right)^{-4/3}}{k + \tau_{\rm LW}^{1/3}\left(\frac{P_s}{\rm 1\,bar}\right)^{2/3}\left(\frac{T_{\rm eq}}{\rm 600\,K}\right)^{-4/3}},
\end{equation}
where $k$ captures all planetary parameters except for those explicitly written out in the equation. We assume a nominal value of $k=2$, which is typical for hot rocky exoplanets, because $k=1.2$, 1.9, and 2.3 for TRAPPIST-1b, GJ 1132 b, and LHS 3844 b, respectively, and $k$ is not sensitive to planetary parameters \citep{koll_scaling_2022}. We further assume $\tau_{\rm LW}=1$.

Figure \ref{fig:tirr_vs_r_fits}b shows the $\mathcal{R}$ profiles of planets with rock vapor atmospheres in equilibrium with MOs ($P_s=P_{\rm eq}$, solid line), moderate MO outgassing ($P_s=10\times P_{\rm eq}$, dashed line), and high MO outgassing ($P_s=100\times P_{\rm eq}$, dotted line), assuming three different Bond albedos: 0 (red), 0.3 (green), and 0.5 (orange).

Even under the assumption of high albedo ($A_B=0.5$) and high MO outgassing ($P_s=100\times P_{\rm eq}$), heat redistribution calculated from the scaling relation is not sufficient to cool the dayside hemisphere to the observed levels. The maximum equilibrium pressure at $\tirr=3500$ K is $\approx2.3$ bar, producing $P_s\approx230$ bar at max under our assumption. Yet, such a thick atmosphere only has a $\varepsilon\approx0.74$. Therefore, our results imply that thick and volatile-rich atmospheres are indeed required to explain the observed thermal deficit of hot rocky exoplanets like 55 Cnc e and TOI-561 b, as claimed by e.g., \cite{hu_secondary_2024} and \cite{teske_thick_2025}. 

Using experiments, \cite{essack_low-albedo_2020} found that surfaces of lava worlds are dark with albedo $\lesssim0.1$, seemingly contradicting our finding that high albedos are needed to explain the thermal deficit of the hottest rocky exoplanets. However, extended Na and K clouds may form in the atmospheres of hot rocky exoplanets \citep{schaefer_chemistry_2009}. \cite{mahapatra_cloud_2017} simulated cloud formation on 55 Cnc e and concluded that mineral clouds (e.g., TiO$_2$, SiO, SiO$_2$, MgO, MgSiO$_3$, Mg$_2$SiO$_4$) can form. Clouds can increase the albedos of rocky exoplanets, offering a possible explanation for the observed thermal deficits.


\subsection{Residual Heating Only Explains Thermal Emission Excess for Very Young Planets} \label{sect:res_residual_heating}

\begin{figure}[t]
    \centering
    \includegraphics[width=\linewidth]{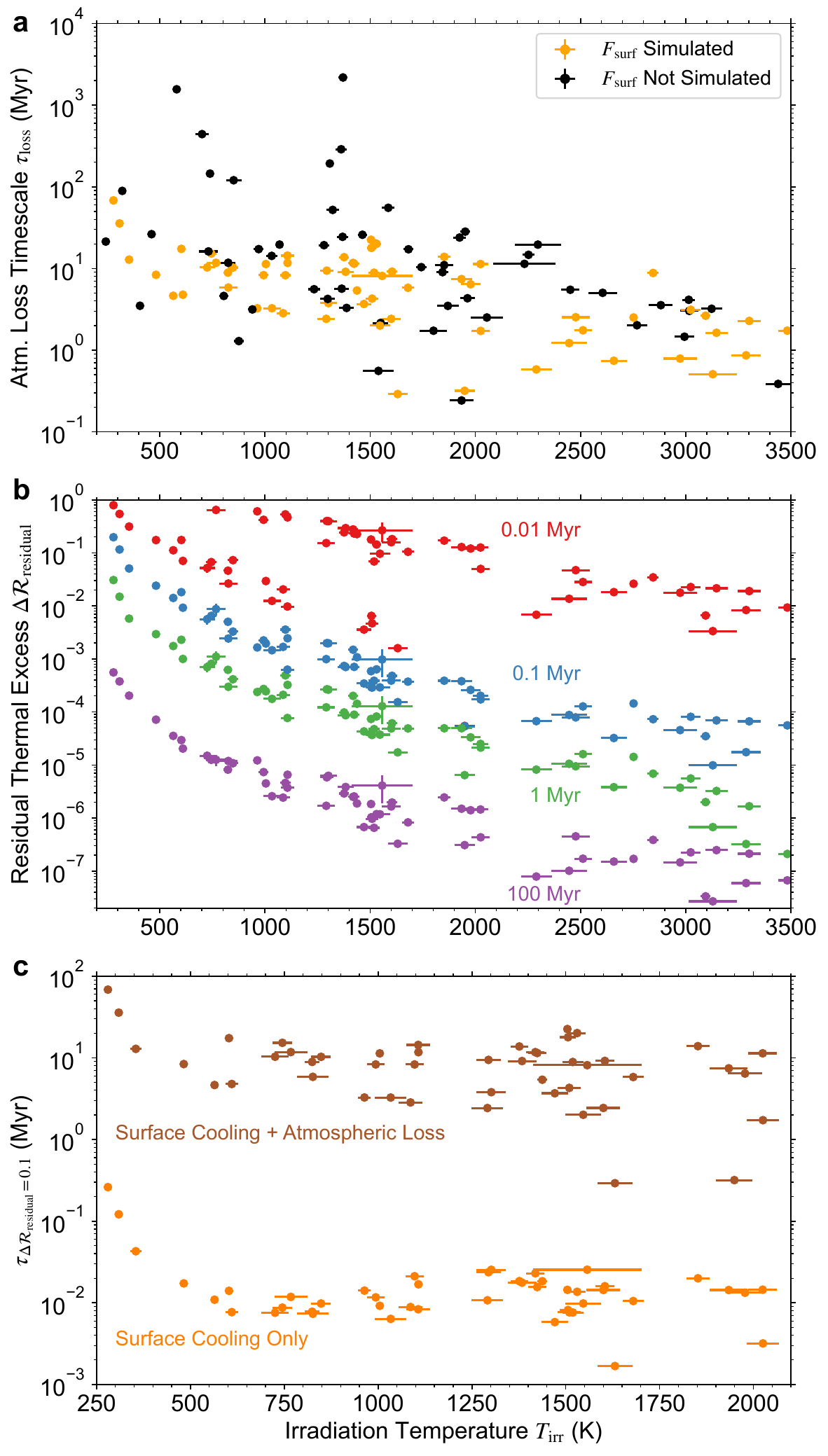}
    \caption{(a) Atmospheric loss timescale $\tau_{\rm loss}$ as a function of $\tirr$ for rocky exoplanets. Planets simulated with \texttt{CMAPPER} are highlighted in orange. (b) Residual thermal excess $\deltar_{\rm residual}$ due to surface heat flux after total atmospheric loss as a function of $\tirr$, at four stages in the planet's evolutionary history: (red) 0.01 Myr after atmospheric loss, (blue) 0.1 Myr, (green) 1 Myr, and (purple) 10 Myr. Note that $\deltar_{\rm residual}$ is negatively correlated with $\tirr$, as opposed to the positive trend observed in Figure \ref{fig:tirr_vs_r_fits}. (c) The time at which $\deltar_{\rm residual}=0.1$, considering (orange) only surface cooling and (brown) atmospheric loss plus surface cooling.}
    \label{fig:r_residual_vs_tirr}
\end{figure}

Our residual heating results (Figure \ref{fig:r_residual_vs_tirr}) depend on two models: the \cite{owen_evaporation_2017} analytical photoevaporation model, and \texttt{CMAPPER} for MO cooling and surface heat flux evolution. Below, we present results for these two mechanisms separately and discuss the implications when they are analyzed jointly. 

\subsubsection{Total Atmospheric Loss is Rapid for Most Rocky Exoplanets}
We find that most rocky exoplanets will lose their H/He envelopes entirely within $\sim100$ Myr, assuming an initial envelope mass fraction of 1\% (Figure \ref{fig:r_residual_vs_tirr}a), in agreement with previous studies showing that photoevaporation is dominated by the the first $\sim100$ Myr when stellar high-energy fluxes are high \citep[e.g.,][]{jackson_coronal_2012, owen_evaporation_2017, rogers_unveiling_2021}. This implies that planetary interiors can be insulated only by envelopes for a short period, after which surface MOs will be exposed and become vulnerable to cooling. Interestingly, because \texttt{CMAPPER} is only valid for a limited parameter space ($0.5\, M_\oplus \leq M_p \leq 8\, M_\oplus$, $0.1<$ CMF $<0.7$, and 255 K $\leq T_{\rm eq} \leq2700$ K), there are several planets with long $\tau_{\rm loss}\gtrsim100$ Myr and even $\gtrsim1000$ Myr that are not simulated with \texttt{CMAPPER} for their surface heat flux evolution. Among these planets are HD 63433 b ($\tau_{\rm loss}\approx2180$ Myr, $M_p=37.3\,M_\oplus$), HD 219134 d ($\tau_{\rm loss}\approx1560$ Myr, $M_p>16.17\,M_\oplus$), TOI-2322 c ($\tau_{\rm loss}\approx440$ Myr, $M_p=18.10\,M_\oplus$), and HD 28109 b ($\tau_{\rm loss}\approx290$ Myr, $M_p=18.496\,M_\oplus$). These planets represent a group of massive super-Earths capable of sustaining envelopes for geologic timescales, yet with bulk densities high enough to be consistent with rocky compositions. They may be ideal targets for detecting residual heat, as their atmospheres can keep the MOs hot for an extended period.

So far, our simulations only consider H/He atmospheres. High mean molecular weight atmospheres, such as N$_2$- or CO$_2$-dominated atmospheres, are more resilient to photoevaporation and may insulate the surface over longer timescales, despite having a weaker greenhouse effect than H/He atmospheres. \cite{kite_exoplanet_2020} studied the retention and revival by volcanic outgassing of secondary atmospheres on rocky exoplanets, concluding that some of the coolest planets in our sample, such as TRAPPIST-1 b and LHS 1140 c, may have residual or revived secondary atmospheres. However, recent modeling work on the retention distance of N$_2$- or CO$_2$-dominated atmospheres \citep{looveren_habitable_2025} concluded that all planets in our sample -- even the coolest TRAPPIST-1 c -- are too close to their host stars for atmospheric retention. \cite{pass_receding_2025} confirmed this result by finding that M-Earths receive $\sim2$--3 times more cumulative XUV flux from their host stars than previously expected, causing the cosmic shoreline \citep{zahnle_cosmic_2017} to recede and making atmospheric retention less likely. Therefore, our conclusion that total atmospheric loss is rapid for rocky exoplanets in our sample is robust, regardless of the atmospheric composition.

\subsubsection{Surface Heat Flux Cannot Explain the Thermal Emission Trend}
We simulate the magma solidification and surface heat flux evolution for 68 rocky exoplanets using \texttt{CMAPPER} to track how their surfaces cool down after total atmospheric loss. Residual thermal excess, $\deltar_{\rm residual}$, at several evolutionary stages is shown as a function of $\tirr$ in Figure \ref{fig:r_residual_vs_tirr}b. We find that surface heat flux cannot explain the observed trend in thermal emission for two reasons. 

The first reason is that rocky planet surfaces cool down too rapidly. Only within the first $\sim0.01$--0.1 Myr after total atmospheric loss can a planet produce enough surface heat flux to generate $\deltar_{\rm residual} \sim 0.1$. At only 10 Myr after atmospheric loss, $\deltar_{\rm residual}$ of all planets decreases to an undetectable level of $\lesssim10^{-3}$.

The second reason is that $\deltar_{\rm residual}$ is negatively correlated with $\tirr$, suggesting that even if a population of planets that recently lost their envelopes is observed, a positive $\deltar_{\rm residual}$ trend is unlikely to be identified. This negative correlation arises because surface heat flux $F_{\rm surf}$ is not a sensitive function of equilibrium temperature \citep{zhang_thermal_2022}, but the insolation flux $F_{\rm insol}$ is. Hence, as the $\tirr$ of a planet increases, the relative contribution from residual heat flux to the overall heat flux rapidly decreases.

\subsubsection{Combined Results: Residual Heating Only Significant for Less than 100 Myr}
Here, we report the results of a joint analysis of atmospheric escape and the evolution of surface heat flux. In Figure \ref{fig:r_residual_vs_tirr}c, we show the timescale for which residual heating produces a detectable level of thermal emission excess ($\deltar_{\rm residual}=0.1$), assuming surface cooling only (orange) and surface cooling after atmospheric loss (brown). In summary, no planet can maintain a high $\deltar_{\rm residual}=0.1$ for more than 100 Myr. Except for two temperate rocky exoplanets with $\tirr\lesssim350$ K, the vast majority of planets have $\tau_{\deltar_{\rm residual}=0.1}\sim10$ Myr. As a result of this short timescale, it is implausible that any thermal emission excess we observe today is produced by residual heat.


\subsection{Tidal Heating is Unlikely to Explain Observed Thermal Emission Excess} \label{sect:res_tidal_heating}

\begin{figure}[t]
    \centering
    \includegraphics[width=\linewidth]{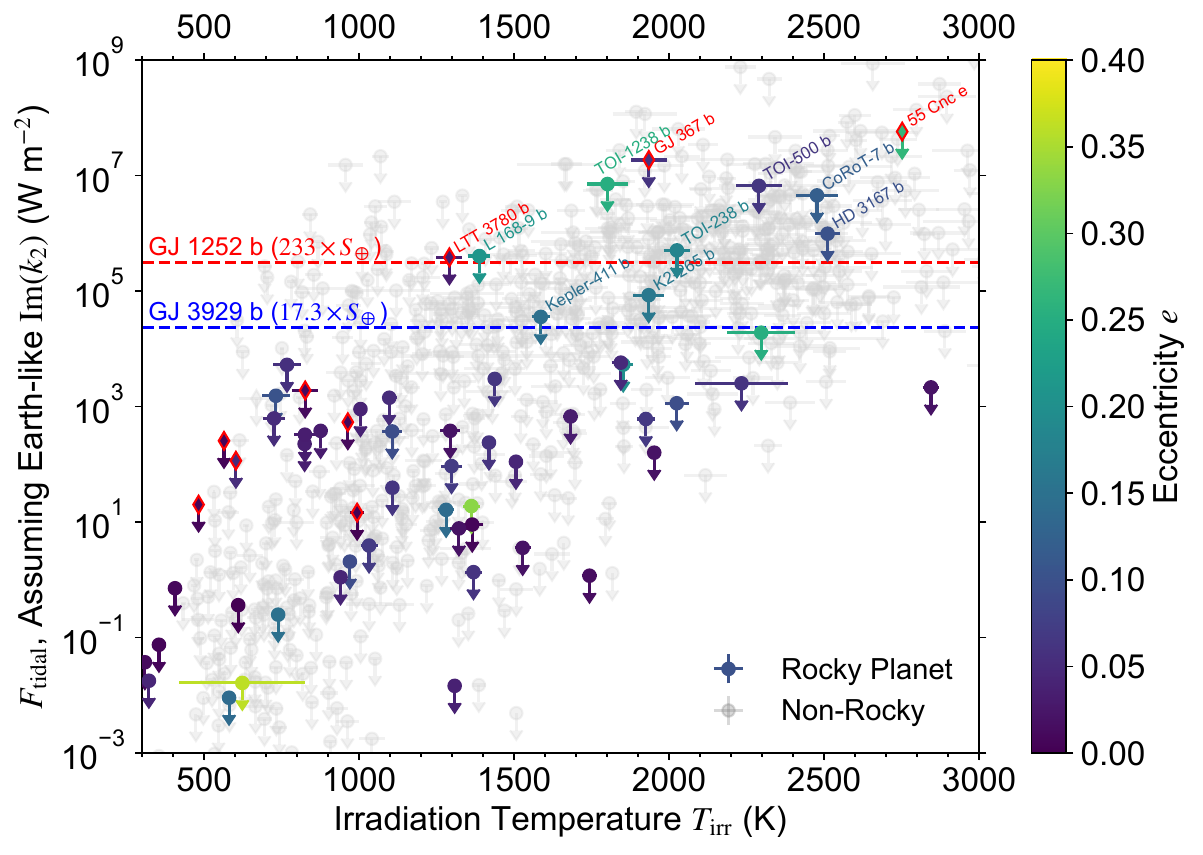}
    \caption{Upper limit on tidal heat flux, $F_{\rm tidal}$, as a function of irradiation temperature, $\tirr$, assuming Earth-like tidal ${\rm Im}(k_2)$. Rocky exoplanets ($p_{\rm rocky}\geq32\%$) are color-coded by their eccentricities, while non-rocky exoplanets are shown in gray. Observed rocky exoplanets (Table \ref{tab:obs_planets_data}) are highlighted with red edges. Planets with $F_{\rm tidal}$ greater than the irradiation of GJ 3929 b ($17.3\times S_\oplus$) are annotated. $F_{\rm tidal}$ are shown as upper limits for clarity because many exoplanets have $e$ within error with, or very close to, zero.}
    \label{fig:tirr_vs_ftidal}
\end{figure}

\begin{figure*}[t]
    \centering
    \includegraphics[width=\textwidth]{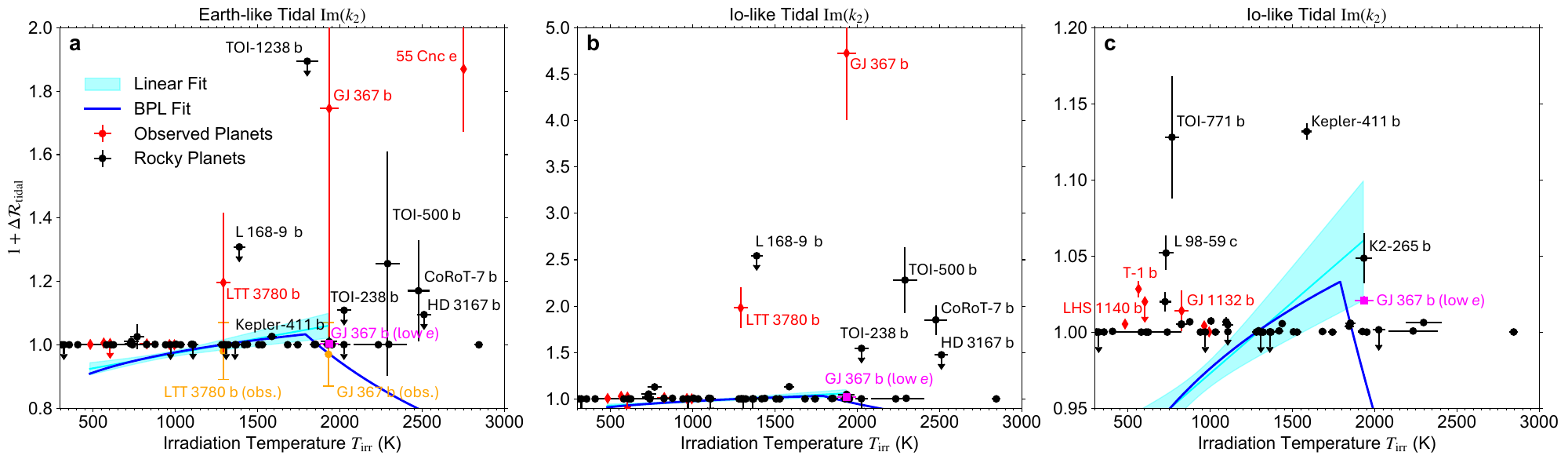}
    \caption{Thermal emission excess due to tidal heating, $1+\deltar_{\rm tidal}$, as a function of irradiation temperature for (a) Earth-like tidal $-{\rm Im}(k_2)$ parameter, and (b) Io-like tidal $-{\rm Im}(k_2)$ parameter. (c) is the same as (b), but zoomed in to the region most relevant for the positive $\mathcal{R}$ trend for M-Earths.}
    \label{fig:R_tidal_3tidalQ}
\end{figure*}

\begin{figure}[t]
    \centering
    \includegraphics[width=\linewidth]{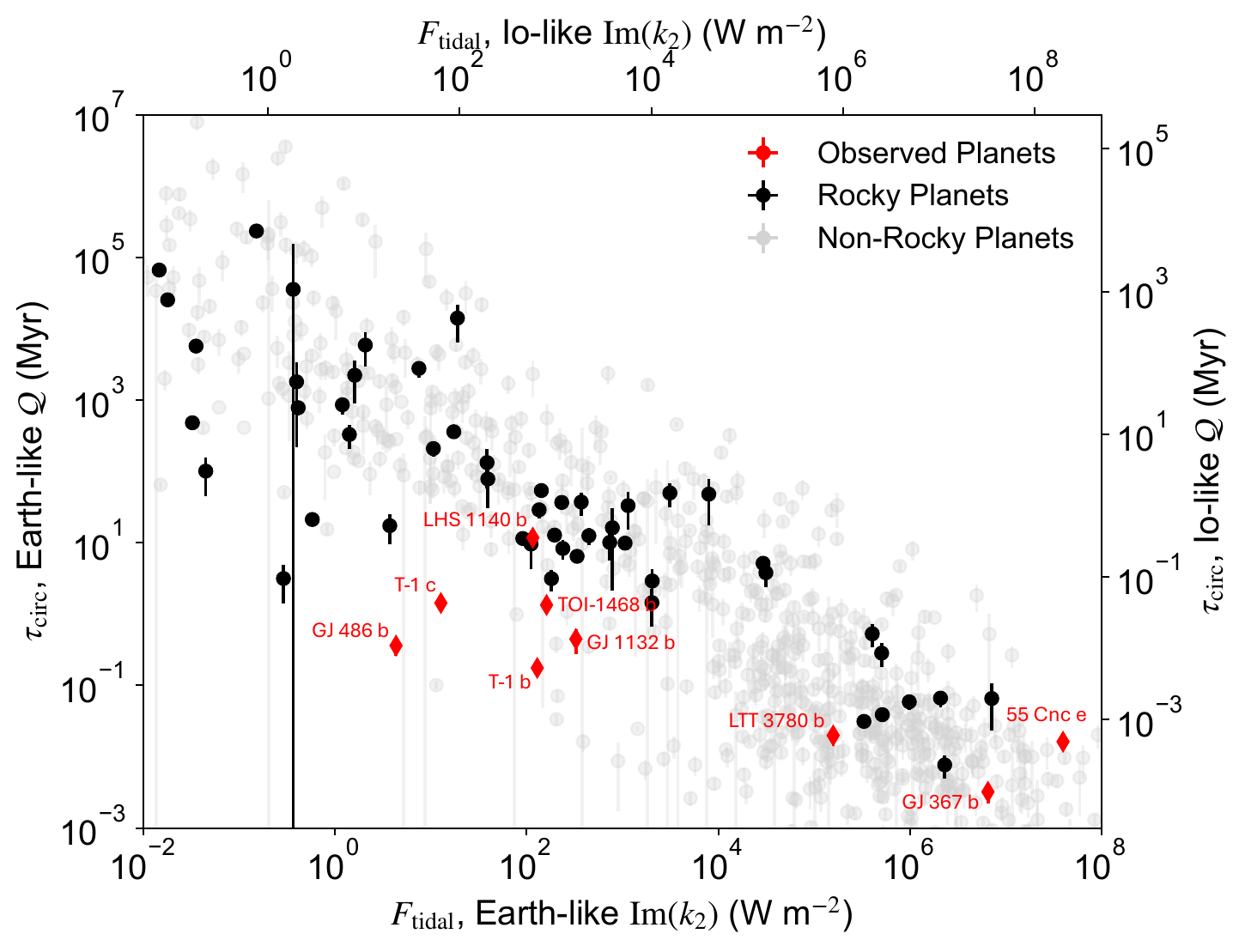}
    \caption{Tidal circularization timescale $\tau_{\rm circ}$ as a function of tidal heat flux $F_{\rm tidal}$. Both Earth-like and Io-like ${\rm Im}(k_2)$ and $\mathcal{Q}$ results are shown. Planets with large tidal $\deltar$ (Figure \ref{fig:R_tidal_3tidalQ}) are also vulnerable to rapid orbital circularization with $\tau_{\rm circ}$ as short as $\lesssim0.1$ Myr. Detecting large $\deltar_{\rm tidal}$ is therefore unlikely, unless high $e$ can be maintained.}
    \label{fig:tau_circ_vs_ftidal}
\end{figure}

We use the fixed-$\mathcal{Q}$ model \citep{driscoll_tidal_2015} to simulate the tidal heat production of known exoplanets with nonzero eccentricities. Even though tidal heating is sometimes invoked to explain the observed thermal emissions of rocky exoplanets \citep[e.g.,][]{zhang_gj_2024, valdes_hot_2025}, we found that on a population level, tidal heating is unlikely to explain the positive thermal emission trend shown in Figure \ref{fig:tirr_vs_r_fits}. 

The tidal heat flux, $F_{\rm tidal}$, of most rocky exoplanets is orders of magnitude lower than their insolation (Figure \ref{fig:tirr_vs_ftidal}), if assuming Earth-like ${\rm Im}(k_2)$. The irradiation of GJ 1252 b ($\tirr=1542$ K), which receives $233\times$ Earth's insolation, $S_\oplus$ \citep{crossfield_gj_2022}, and GJ 3929 b ($\tirr=822$ K), which receives $17.3\times S_\oplus$ \citep{beard_gj_2022}, representing the approximate irradiation range of hot rocky exoplanets, are shown for comparison. Among all rocky exoplanets, only 11 have tidal heating rates that are comparable to or exceed the irradiation of GJ 3929 b, three of which have dayside emission observations (LTT 3780 b, GJ 367 b, and 55 Cnc e). If rocky exoplanets indeed have Earth-like tidal dissipation, tidal heating should be insufficient to explain the observed positive $\mathcal{R}$ trend.

Indeed, as shown in Figure \ref{fig:R_tidal_3tidalQ}a, $\deltar_{\rm tidal}$ of most rocky exoplanets is too small to explain the positive $\mathcal{R}(\tirr)$ trend for M-Earths. Only five M-Earths -- LTT 3780 b, L 168-9 b, Kepler-411 b, TOI-1238 b, and GJ 367 b -- have high tidal heat fluxes that can explain the positive trend. The other planets with large $\deltar_{\rm tidal}$ have $\tirr$ that exceed the $\tirrroche$ of M dwarf exoplanets, so unless the positive trend can be extrapolated to some planets orbiting hotter stars, they do not explain the observed thermal excess. The observed $\mathcal{R}=0.98\pm0.09$ of LTT 3780 b \citep{allen_hot_2025} is consistent with a hot bare rock, despite its high calculated $\deltar_{\rm tidal}$, implying that the planet is not significantly heated by tides, possibly as a result of circularization. The predicted $\deltar_{\rm tidal}$ for GJ 367 b may be an overestimation, because it adopted a high eccentricity value of $e=0.060^{+0.070}_{-0.040}$ reported by \cite{goffo_company_2023}. However, a more recent JWST observation suggested that GJ 367 b has a nearly circular orbit with $e\cos\omega = 0.0027\pm0.0008$ \citep{zhang_gj_2024}. If we assume $e=e\cos\omega$, the tidal contribution to $\mathcal{R}$ of GJ 367 b will drop significantly from $\deltar = 0.75\pm0.72$ to $\deltar = 0.0042\pm0.0028$ (Figure \ref{fig:R_tidal_3tidalQ}a, magenta square marker).

We further assume that rocky exoplanets may have tidal responses much stronger than Earth and similar to that of Io (Figure \ref{fig:R_tidal_3tidalQ}b,c). Given the Io-like $-{\rm Im}(k_2)=0.015$, which is five times stronger than that of the Earth, some planets can in theory have extremely high tidal thermal excess ($\deltar_{\rm tidal}\gtrsim0.5$, Figure \ref{fig:R_tidal_3tidalQ}b). However, the simulated $\deltar_{\rm tidal}$ results do not agree with the linear or BPL fits (Figure \ref{fig:R_tidal_3tidalQ}c). Only GJ 367 b (assuming low $e$ measured by \citealt{zhang_gj_2024}), K2-265 b, and Kepler-411 b (assuming Earth-like $-{\rm Im}(k_2)$, see Figure \ref{fig:R_tidal_3tidalQ}a) follow the fitted positive $\deltar_{\rm tidal}$ trend. Several planets with $\tirr\lesssim1000$ K can potentially have strong tidal thermal excess, including LHS 1140 c, where its positive large $\deltar_{\rm tidal}$ possibly explains its $\mathcal{R}=1.04\pm0.08$ that is slightly above 1 \citep{fortune_hot_2025}. TRAPPIST-1 b and GJ 1132 b can potentially have detectable $\deltar_{\rm tidal}$, but their observed $\mathcal{R}$ are below 1. In summary, tidal heating does not provide satisfactory explanations for the observed thermal excess of rocky exoplanets. Nevertheless, some rocky exoplanets, including TOI-1238 b, L 168-9 b, and TOI-500 b, may still be interesting targets for the search for tidally generated heat fluxes, if they indeed have high eccentricities. TOI-500 b, for example, has been observed by MIRI/LRS (GO 4818, PI: Megan Weiner Mansfield).

Short tidal circularization timescales make it even less likely that tidal heating is responsible for the thermal emission excess. Figure \ref{fig:tau_circ_vs_ftidal} shows $\tau_{\rm circ}$ as a function of $F_{\rm tidal}$ for both Earth-like and Io-like $-{\rm Im}(k_2)$. All observed rocky exoplanets have $\tau_{\rm circ}\lesssim100$ Myr, much shorter than their ages. Furthermore, $\tau_{\rm circ}$ rapidly decreases as $F_{\rm tidal}$ increases, so planets like LTT 3780 b and GJ 367 b, which are predicted to have high $\deltar$, can be circularized within 0.1 Myr unless mechanisms sustaining their eccentricities exist. Because $\tau_{\rm circ} \propto \mathcal{Q}$, and Io's $\mathcal{Q}\approx3$ is much smaller than that of the Earth ($\mathcal{Q}\approx100$), tidal circularization is much more rapid for planets with Io-like tidal properties, making observing high tidal heat fluxes at present day unlikely.


\subsection{Induction Heating is Insufficient to Explain Observed Thermal Emission Excess} \label{sect:res_induction_heating}

\begin{figure}[t]
    \centering
    \includegraphics[width=\linewidth]{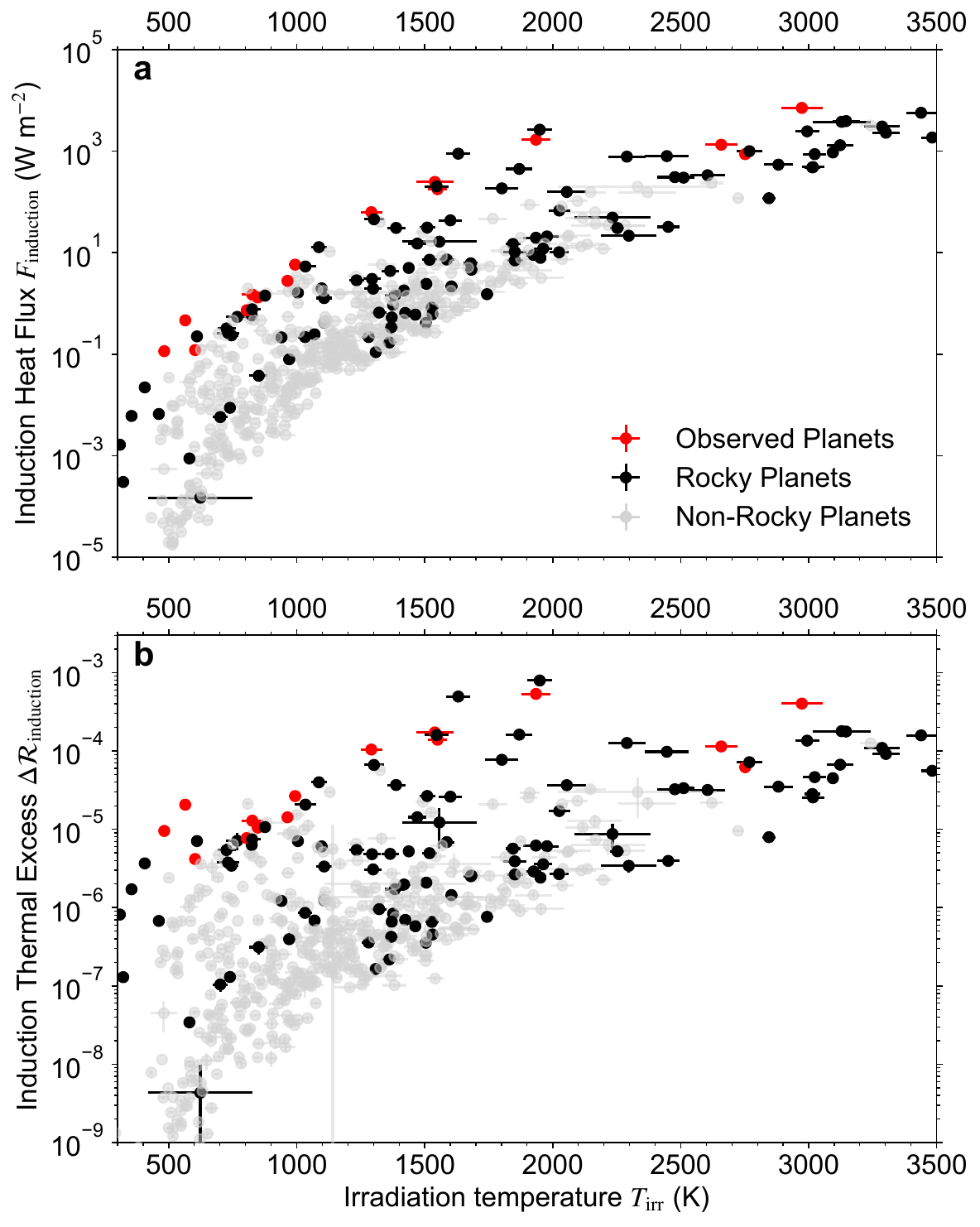}
    \caption{(a) Induction heat flux $F_{\rm induction}$ as a function of $\tirr$. (b) Thermal emission excess produced by induction heat flux $\deltar_{\rm induction}$ as a function of $\tirr$. Rocky exoplanets are shown in black; observed rocky exoplanets are highlighted in red; all other planets are shown in gray. Even though some planets may have high induction heating on the order of $\sim10^3$ W m$^{-2}$, the relative contribution to dayside emission is too small ($\deltar<10^{-3}$ for all planets). Induction heating, therefore, cannot explain the observed thermal excess.}
    \label{fig:finduction_vs_tirr}
\end{figure}

The thermal emission excess caused by induction heating for all rocky exoplanets is orders of magnitude smaller than the observed thermal excess. The absolute heat flux due to induction, $F_{\rm induction}$, and the relative contribution to the measured dayside temperature, $\deltar_{\rm induction}$, are shown in Figure \ref{fig:finduction_vs_tirr}. Even though some close-in planets may experience a strong stellar magnetic field that can potentially generate an $F_{\rm induction}$ as strong as $\gtrsim10^3$ W m$^{-2}$, comparable to Earth's insolation, the relative $\deltar_{\rm induction}$ is dwarfed by the extreme irradiation these planets receive. When compared to $F_{\rm insol}$, all planets have $\deltar_{\rm induction} < 10^{-3}$, which is too small to explain planets with $\deltar\sim0.1$ and the positive trend in $\mathcal{R}$ (Figure \ref{fig:tirr_vs_r_fits}). Among the observed exoplanets, GJ 367 b ($\deltar_{\rm induction}=5.35\times10^{-4}$) and K2-141 b ($\deltar_{\rm induction}=4.02\times10^{-4}$) have the highest induction heating potentials, while TOI-6255 b ($\deltar_{\rm induction}=7.93\times10^{-4}$) and TOI-6324 b ($\deltar_{\rm induction}=4.94\times10^{-4}$) ranked top among the other rocky exoplanets.

Our finding that induction heating cannot explain the observed thermal excess is robust despite the simplifications we make to the analytical model, because we choose parameters that place upper limits on induction heat flux. Our assumption that orbital inclination equals 90\textdegree\ is an overestimate for most planets, because planets with low obliquities around zero are more common than those with obliquities around 90\textdegree\ \citep{frazier_neid_2023}. One complication is that there may be an angle between the rotation axis and dipole axis of the host star \citep{kislyakova_magma_2017}. Nevertheless, 90\textdegree\ provides a firm upper bound. Decreasing the obliquity will lead to a $\sim1$--2 orders of magnitude drop in $Q_{\rm induction}$, when all other assumptions are held equal \citep{kislyakova_effective_2018}. The stellar field we assume (7.3 kG) is also one of the strongest fields among active M dwarfs \citep{shulyak_magnetic_2019}, placing another firm upper bound.

A potential caveat is that we may underestimate the frequency of magnetic field variation, $\omega$. The frequency is related to both the host star's rotation frequency ($\omega_{\rm *}$) and the planet's orbital frequency ($\omega_p$)
\begin{equation}
    \omega = |\omega_* - \omega_p|,
\end{equation}
if the planet is on a prograde orbit \citep{kislyakova_magma_2017}. If the planet is on a retrograde orbit, one simply flips the sign to $\omega = |\omega_* + \omega_p|$. Because the rotation rates of the majority of exoplanet host stars are unconstrained, we assume $\omega = \omega_p = \sqrt{GM_*/a^3}$ for most exoplanets. This introduces some uncertainty to our results, because $Q_{\rm induction} \propto \omega^2$. However, even though some active M dwarfs have rapid rotations with periods $<1$ day \citep{shulyak_magnetic_2019}, most exoplanet host stars rotate much more slowly. Among all observed systems, TRAPPIST-1 has the most rapid rotation ($P_{\rm rot}=1.4$ day, \citealt{gillon_temperate_2016}), while other stars have rotation periods on the order of tens of days (e.g., GJ 367 has $P_{\rm rot}=51.30\pm0.13$ days, \citealt{goffo_company_2023}), or even $>100$ days (e.g., GJ 3929 has $P_{\rm rot}=122\pm13$ days, \citealt{kemmer_discovery_2022}). Such slow rotation rates are insufficient to increase $\omega$ significantly, given that highly irradiated planets generally have much shorter ($\lesssim1$ day) orbital periods.


\subsection{Combined Results: Internal Mechanisms Do Not Produce Significant Thermal Emission Excess} \label{sect:res_combined}

\begin{figure*}[t]
    \centering
    \includegraphics[width=\textwidth]{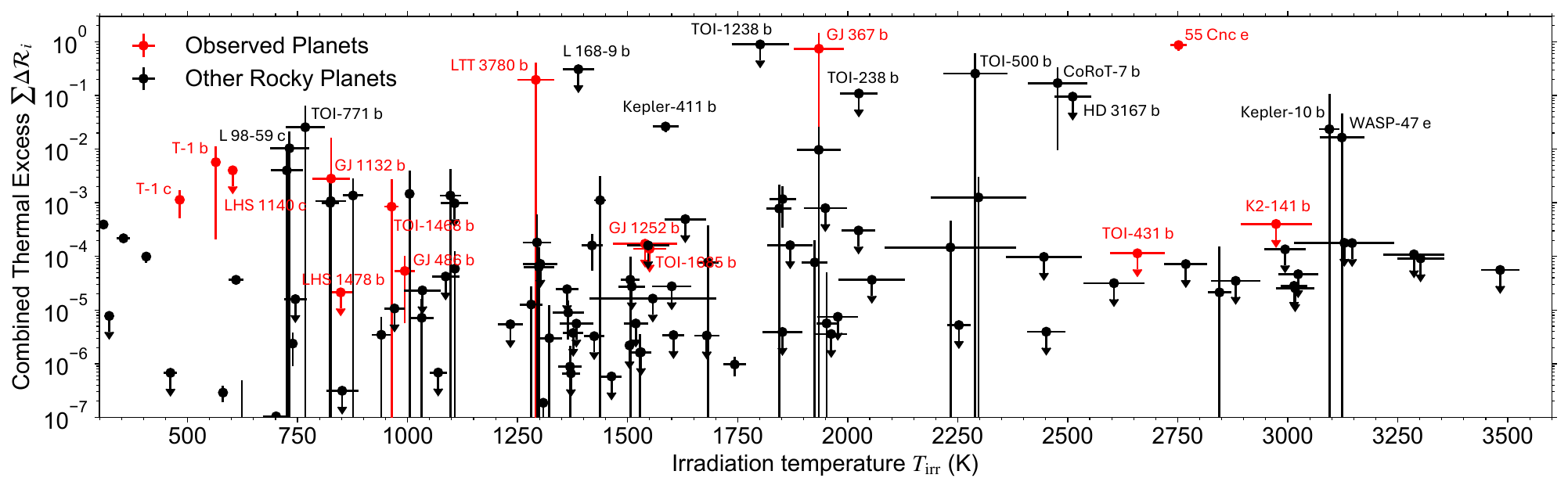}
    \caption{Combined thermal emission excess $\Delta\mathcal{R}_{\rm total} = \Delta\mathcal{R}_{\rm residual}+\Delta\mathcal{R}_{\rm tidal}+\Delta\mathcal{R}_{\rm indcution}$ as a function of $\tirr$. Observed rocky exoplanets are highlighted in red, while all other rocky exoplanets are shown in black. The error bars sum the uncertainties in $\Delta\mathcal{R}_i$. Arrows indicate upper limits when uncertainties are unavailable. Observed exoplanets and planets with $\Delta\mathcal{R}_{\rm total} > 0.01$ are annotated, where the latter are interesting targets for future characterizations.}
    \label{fig:Rcombined_vs_Tirr}
\end{figure*}

Given the results for residual heating, tidal heating, and induction heating above, we may calculate the combined thermal excess potential $\Delta\mathcal{R}_{\rm total}$ using Equation (\ref{eq:total_r}). We assume the following typical conditions: $\Delta\mathcal{R}_{\rm residual}$ assumes 100 Myr after total atmospheric loss, $\Delta\mathcal{R}_{\rm tidal}$ assumes Earth-like $-{\rm Im}(k_2)$, and $\Delta\mathcal{R}_{\rm induction}$ assumes the default parameters that produce results shown in Figure \ref{fig:finduction_vs_tirr}.

The combined thermal emission excess, $\Delta\mathcal{R}_{\rm total}$, is shown in Figure \ref{fig:Rcombined_vs_Tirr} as a function of irradiation temperature $\tirr$. As discussed above, surface flux due to residual heat is negligible at a reasonable planetary age ($\gtrsim100$ Myr), and induction heating is also insignificant ($\Delta\mathcal{R}_{\rm induction} < 10^{-3}$ for all planets). Therefore, the combined $\Delta\mathcal{R}_{\rm total}$ is primarily dominated by tidal heat flux. Several observed exoplanets, including LTT 3780 b, GJ 367 b, and 55 Cnc e, have high $\Delta\mathcal{R}_{\rm total}$, but with some caveats that were discussed in Section \ref{sect:res_tidal_heating}.

In summary, we do not identify any planetary internal heating mechanism that can reliably produce the observed thermal emission excess for rocky exoplanets on the order of $\Delta\mathcal{R}\sim0.1$, nor can we identify any internal heating mechanism that produces a positive $\mathcal{R}$ trend with $\tirr$. The observed excess and trend are therefore likely due to other physical mechanisms not modeled in this work, instrumental noises, or stellar interferences. In the following section, we will discuss other possible explanations of thermal excess.

\section{Discussion} \label{sect:discussion}

Several physical mechanisms not explored above may explain the observed excess in thermal emission. These mechanisms include stellar contamination, planetary surface albedo and roughness effects, geometric effect, and annihilation heating due to dark matter. Below, we will discuss each of these effects in Section \ref{sect:dis_stellar_conta}--\ref{sect:dis_dark_matter}. Finally, we discuss the prospects of rocky exoplanet emission observations in the era of the Nancy Grace Roman Space Telescope in Section \ref{sect:dis_roman}. Instrumental effects and differences across data reduction pipelines may also bias results and lead to overestimation of $\mathcal{R}$, but a detailed discussion of these effects is beyond the scope of this work. In addition, transient processes can lead to time-variable, non-equilibrium heating. Such processes include heating by giant impacts, induction heating by coronal mass ejections, or tidal heating by recent orbital migrations. These mechanisms are left for future studies.

\subsection{Stellar Contamination} \label{sect:dis_stellar_conta}

To date, all rocky exoplanets found to have thermal emission excesses orbit M dwarfs. M dwarfs are known to be active \citep{gunther_stellar_2020}. Frequent flares or surface spots and faculae may interfere with observations, contaminating planetary signals and leading to overestimation or underestimation in dayside emission temperatures.

In recent years, stellar contamination has been identified as a fundamental challenge in precise exoplanet characterization via transmission spectroscopy \citep[e.g.,][]{rackham_transit_2018, rackham_effect_2023}. The authors showed that unocculted starspots and faculae can lead to systematic biases in measured planetary radii (and hence density), and can shift apparent transit depths by amounts comparable to or exceeding the signatures of small-planet atmospheres. 

Similarly, stellar contamination limits the precision of exoplanet characterization via emission spectroscopy. \cite{fauchez_stellar_2025} pointed out that imperfect knowledge of stellar spectra can bias the derived planetary emission flux by $\gtrsim20\%$ for mid- to late-M dwarfs. Indeed, \cite{coy_population-level_2025} discussed the difference in $\mathcal{R}$ derived from the SPHINX model, the PHOENIX model, and observed stellar spectra. For GJ 367 b, $\mathcal{R}=1.074\pm0.047$ was reported using the SPHINX model, which is $1\,\sigma$ higher than $\mathcal{R}=0.97\pm0.10$ reported in the original data paper using the stellar spectrum observed by MIRI \citep{zhang_gj_2024}.

Therefore, reliable stellar models will significantly benefit our interpretation of rocky exoplanet thermal emissions. The maximum $\Delta\mathcal{R}_{\rm total}$ among all rocky exoplanets is on the order of $\sim0.1$, if ignoring the few planets with $\Delta\mathcal{R}_{\rm total}\sim1$ upper bounds primarily due to large eccentricity values (Figure \ref{fig:Rcombined_vs_Tirr}). This level of thermal emission excess is comparable to the uncertainty in $\mathcal{R}$ introduced by stellar models. Future emission observations should include some mitigation strategy, such as collecting MIRI stellar spectra for each secondary eclipse visit \citep{fauchez_stellar_2025}.

\subsection{Surface Albedo and Roughness} \label{sect:dis_albedo_roughness}

In this work, we focus on internal planetary effects that can generate excessive thermal emission while ignoring surface effects. Surface albedo and the ``thermal beaming'' effect that is related to surface roughness, however, can change the dayside emission temperature of a rocky exoplanet.

Surface albedo has a direct impact on the measured $\tday$. Long-term space weathering of bare rocks changes surface albedo by reducing grain size and generating dark fine-grained metallic Fe \citep[see e.g.,][and references therein]{lyu_super-earth_2024, coy_population-level_2025}, and higher $\tirr$ implies more intense and frequent space weathering. Therefore, the positive $\mathcal{R}$ trend can be partially explained by a trend in surface albedo, for the segment where $\mathcal{R}\leq1$. \cite{essack_low-albedo_2020} experimentally studied the surface albedo of lava worlds, and \cite{hammond_reliable_2025} simulated dayside emissions of various surface types measured by JWST/MIRI in the F1500W and F1280W filters. Future observations of rocky exoplanets with $\tirr\lesssim1300$ K (roughly where the fitted trend intersects with $\mathcal{R}=1$) and experimental or modeling studies of the albedo of realistic planetary surface materials will inform whether albedo effects can partially explain the thermal emission trend.

Thermal beaming provides a possible mechanism to produce thermal excess. Thermal beaming occurs when the planetary surface is rough, such that facets tilted towards the observer near secondary eclipse are also tilted towards the star, and therefore appear to be hotter than a smooth spherical surface \citep[e.g.,][]{wohlfarth_advanced_2023}. Assuming a Moon-like roughness, \cite{coy_population-level_2025} found that thermal beaming may lead to a $\sim2$--5\% increase in $\tday$. However, how surface roughness correlates with irradiation and space weathering at different orbital distances remains to be investigated.

Furthermore, porous and rough regolith surfaces produced by space weathering can alter the thermal emission features of airless rocky planets, via solid-state greenhouse or anti-greenhouse effects \citep{lyu_impact_2025}. The authors found that subsurface temperature gradients on the order of several hundred kelvin may be present, potentially leading to super-blackbody emissions at long ($>15$ $\mu$m) wavelengths for some surface materials.

\subsection{Geometric Effect} \label{sect:dis_geometric_eff}

When a planet is sufficiently close to its host star, the typical assumption that incoming starlight arrives as plane-parallel rays breaks down \citep[e.g.,][]{carter_analysis_2022, perera_improving_2023, carter_hyper_2024, sadh_importance_2026}. The ``nightside'' of such a planet is partially illuminated by light rays from the edge of the host star, making the overall illuminated area of the planet exceed 50\%. Such a ``hyper illumination'' effect can change the planet's thermal state, leading to elevated dayside and nightside brightness temperatures. The impact of this geometric effect on secondary eclipse and phase curve observations requires further investigations.

\subsection{Dark Matter} \label{sect:dis_dark_matter}

If dark matter has a nonzero scattering cross section with ordinary matter, massive celestial bodies, including planets, should be able to capture dark matter. Captured dark matter will concentrate in the planet's core, annihilate, and produce heat that may be detectable by remote observers in the infrared, making exoplanets possible probes of dark matter \citep[e.g.,][]{leane_exoplanets_2021, croon_dark_2024}. Recently, \cite{cappiello_can_2025} studied the flow of heat generated by dark matter annihilation within rocky planets and discussed whether dark matter can melt Earth's core. Given suitable thermal conductivity and dark matter scattering cross section, dark matter annihilation heating in the core may produce a nonzero surface thermal emission excess, $\deltar_{\rm DM}$.

Assuming a large scattering cross section of $1.78\times10^{-37}$ cm$^2$, \cite{cappiello_can_2025} predicted that heat flow generated by dark matter through a spherical shell of radius 1000 km can be as high as $\sim20$ TW if the planet's age $\gtrsim1$ Gyr. Assuming efficient heat transport from $R=1000$ km to the planet's surface with $R=R_\oplus$, dark matter will generate a surface heat flux of $\approx0.04$ W m$^{-2}$. This heat flux is too small to be detectable when compared to insolation flux on the order of $>10^3$ W m$^{-2}$ of rocky exoplanets. Nevertheless, internal melting due to dark matter can have significant implications for magnetic field generation on super-Earths, especially in old systems and those near the Galactic Center, where dark matter density is expected to be higher.

\subsection{Rocky Exoplanets Emission Observations in the Era of Roman} \label{sect:dis_roman}

The Nancy Grace Roman Space Telescope (hereafter Roman; \citealt{spergel_wide-field_2015}) will transform our understanding of rocky exoplanets. Expected to launch in Fall 2026, Roman will conduct three core community surveys using its Wide Field Instrument. In particular, the Galactic Bulge Time Domain Survey (GBTDS) will find $\sim$60,000 to $\sim$200,000 transiting exoplanets, increasing the number of known exoplanets by over an order of magnitude \citep{wilson_transiting_2023}. \cite{tamburo_predicting_2023} predicted that the GBTDS will find $\sim1300$ small ($<4\,R_\oplus$) transiting exoplanets around early-to-mid-M dwarfs. In addition, many transiting exoplanets detected by Roman will have short orbital periods and are therefore prone to experiencing tidal decay. The GBTDS is expected to detect $\sim5$--10 orbital decay events \citep{carden_short_2025}, thereby providing peripheral constraints on the likelihood that tidal heating contributes to the observed thermal emission excess.

\section{Conclusions} \label{sect:conclusion}

Some rocky exoplanets exhibit puzzling thermal emission excesses: their measured dayside brightness temperatures exceed the theoretical maximum. In this work, we investigate whether planetary internal processes, including residual, tidal, and induction heating, can account for the observed thermal emission excesses. Our main conclusion is that none of the simulated planetary internal processes can explain the excessive heat of those planets with temperature scaling factor $\mathcal{R}>1$, nor can they explain the tentative positive trend of $\mathcal{R}$ as a function of irradiation temperature $\tirr$. The combined thermal emission potential, $\Delta\mathcal{R}_{\rm total}$, of most rocky exoplanets is below 0.1 or even 0.01, making thermal excess too weak to be detectable (Figure \ref{fig:Rcombined_vs_Tirr}). Below, we summarize the results of each analysis or simulation.

\begin{enumerate}
    \item We identify a tentative positive $\mathcal{R}(\tirr)$ trend for M-Earths, in agreement with the previous work by \cite{coy_population-level_2025}. Using a linear fit, we find that the intercept is $A=0.879\pm0.036$, and the slope is $B=(9.372\pm3.775)\times10^{-5}$. Leaving one planet out will lead to a $\pm2\times10^{-5}$ change in the slope (Appendix \ref{sect:app_robustness_of_trend}).
    \item We identify a tentative negative $\mathcal{R}(\tirr)$ trend for rocky exoplanets orbiting K-type or hotter stars, using a broken power-law fit. An analytical atmospheric heat redistribution efficiency model reveals that, to explain their low measured $\mathcal{R}$, the hottest rocky exoplanets need high albedos or thick atmospheres with surface pressures $\gtrsim100$ times the equilibrium pressure of a vapor atmosphere generated by surface magma.
    \item We simulate the residual heat evolution of rocky exoplanets using an analytical atmospheric loss model and a 1D rocky planet thermal evolution model. We found that the residual heating potential $\Delta\mathcal{R}_{\rm residual}$ of all rocky exoplanets rapidly drops below 0.1 within 100 Myr after formation. Hence, residual heating cannot explain the observed thermal excess, unless evidence of very recent atmospheric loss is found.
    \item We simulate the tidal heating of all rocky exoplanets using the fixed-$\mathcal{Q}$ method assuming both Earth-like and Io-like tidal $\mathcal{Q}$ parameters, and found that $\Delta\mathcal{R}_{\rm tidal}$ of rocky exoplanets do not follow the positive $\mathcal{R}(\tirr)$ trend. Some planets may experience intense tidal heating, though their eccentricities are uncertain and their tidal circularization timescales are very short.
    \item We simulate the induction heating of all rocky exoplanets, assuming a uniform planetary conductivity profile and a strong stellar magnetic field (7.3 kG). The thermal excesses produced by induction heating, $\Delta\mathcal{R}_{\rm induction}$, are smaller than $10^{-3}$ for all rocky exoplanets and are therefore too small to be detectable.
\end{enumerate}

In addition, we identify some rocky exoplanets with high thermal excess potentials that are intriguing targets for future characterization. TOI-6255 b and TOI-6324 b are among the closest-in M-Earths on the edge of falling into the Roche distance, where the former is the target of a planned JWST/MIRI phase curve observation (GO 8864, PI: Lisa Dang). HD 63433 b, HD 219134 d, TOI-2322 c, and HD 28109 b are potential targets for the search for residual heat because of their long atmospheric loss timescales. TOI-1238 b, L 168-9 b, and TOI-500 b potentially have high tidal heat fluxes, if their high eccentricities are maintained.

Our results imply that if the observed thermal emission excesses of rocky exoplanets are indeed astrophysical rather than instrumental, they must arise from mechanisms not simulated in this work. Possible mechanisms include stellar contamination, surface albedo and roughness effects, and even annihilation heating of dark matter. The JWST-HST joint Rocky Worlds DDT Program will provide additional data points that may elucidate the thermal excess problem and may also detect atmospheres on rocky exoplanets. Expected to launch in October 2026, Roman will significantly increase the number of known exoplanets and provide deeper insights into their thermal emission.

\begin{acknowledgments}
Z.L. and T.D. acknowledge support from the McDonnell Center for the Space Sciences at Washington University in St. Louis. This research has made use of the NASA Exoplanet Archive, which is operated by the California Institute of Technology, under contract with the National Aeronautics and Space Administration under the Exoplanet Exploration Program.
\end{acknowledgments}

\facilities{NASA Exoplanet Archive \citep{christiansen_nasa_2025}}

\software{\texttt{CORGI} \citep{lin_interior_2025}, \texttt{CMAPPER} (\href{https://github.com/zhangjis/CMAPPER_rock}{https://github.com/zhangjis/CMAPPER\_rock}), \texttt{Matplotlib} \citep{Hunter_2007_matplotlib}, \texttt{NumPy} \citep{harris2020array}, \texttt{SciPy} \citep{2020SciPy-NMeth}}

\appendix

\section{Robustness of $\mathcal{R}$ Trends} \label{sect:app_robustness_of_trend}

\begin{figure}[t]
    \centering
    \includegraphics[width=0.9\linewidth]{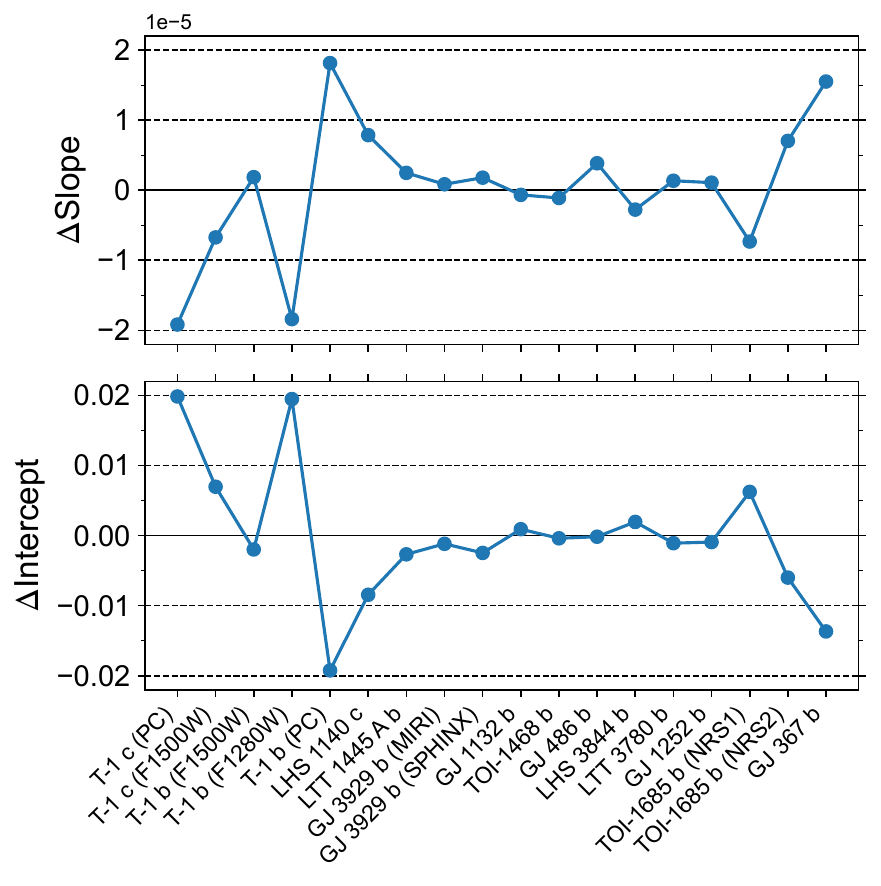}
    \caption{LOO test results for linear fit to the $\mathcal{R}$ as a function of $\tirr$ of M-Earths.}
    \label{fig:LOO_results_linear}
\end{figure}

\begin{figure}[t]
    \centering
    \includegraphics[width=\linewidth]{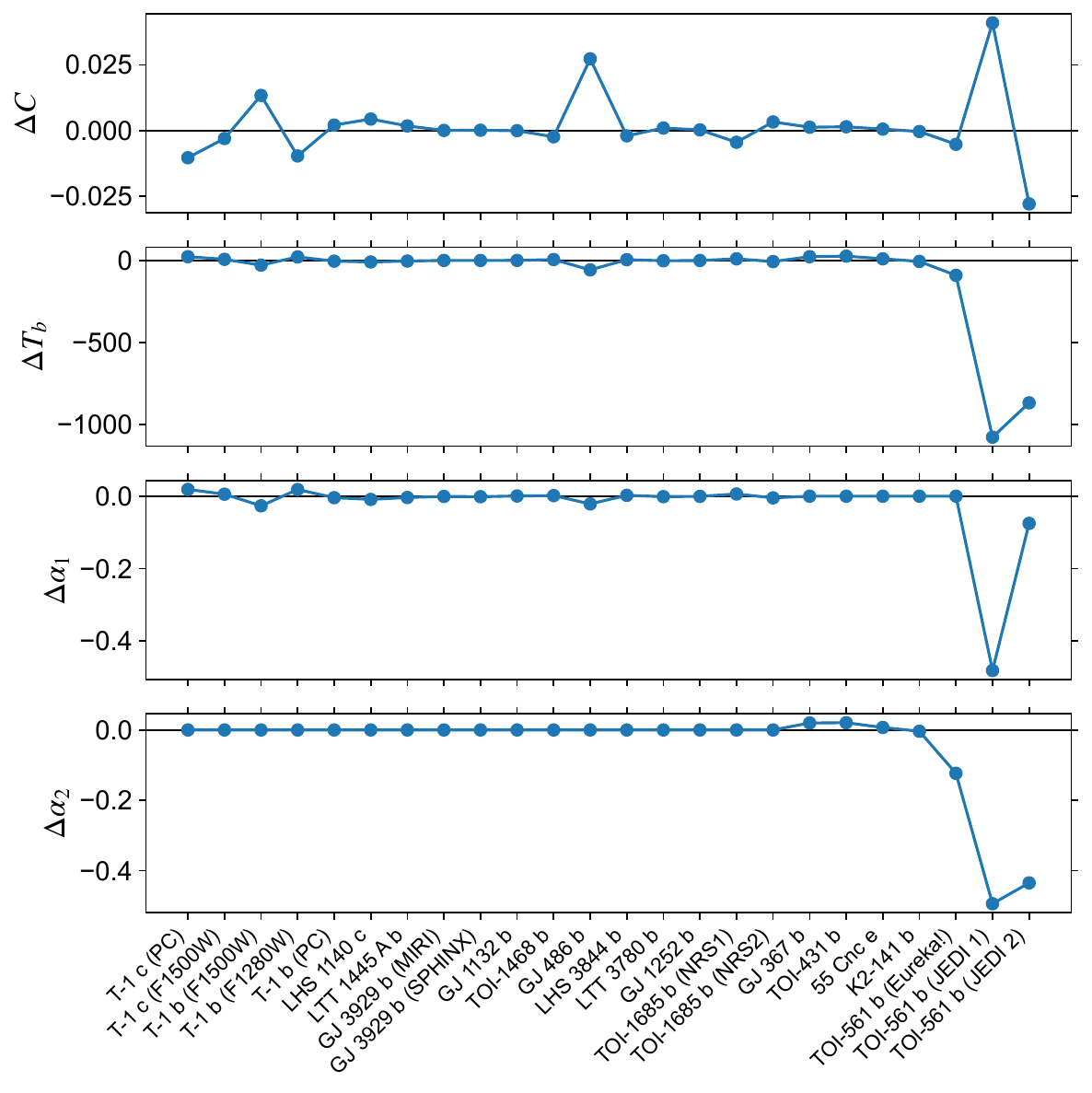}
    \caption{LOO test results for BPL fit to the $\mathcal{R}$ as a function of $\tirr$ of all observed rocky exoplanets.}
    \label{fig:LOO_results_BPL}
\end{figure}

Here, to test the robustness of the positive and negative $\mathcal{R}$ as a function of $\tirr$ trends identified in Section \ref{sect:results}, we perform leave-one-out (LOO) tests. Each time, we remove one planet and perform a linear or BPL fit on the remaining subset, and track the changes in slope $B$ and intercept $A$ (for linear fit), and the changes in $C$, $T_b$, $\alpha_1$, and $\alpha_2$ (for BPL fit). Figure \ref{fig:LOO_results_linear} shows the LOO test results for the linear fit, while Figure \ref{fig:LOO_results_BPL} shows the results for the BPL fit.

Given the small numbers in both the M-Earth and all observed rocky exoplanet datasets, it is not surprising that some planets disproportionately drive changes in the fitted trend. For M-Earths, the linear fit slope and intercept are strongly affected by TRAPPIST-1 b, TRAPPIST-1 c, and GJ 367 b -- the coolest or hottest M-Earths. For BPL fits on all observed rocky exoplanets, the break point temperature $T_b$ and the powers $\alpha_1$ and $\alpha_2$ are most strongly affected by TOI-561 b, the hottest rocky exoplanet observed to date. In addition, TRAPPIST-1 b and GJ 486 b have clear impacts on the constant $C$, likely due to their small $\mathcal{R}$ uncertainties. To provide more robust constraints on the $\mathcal{R}$ trends, future observations should focus on emission of the most temperate ($\tirr\sim500$ K) and hottest ($\tirr\gtrsim3000$ K) exoplanets, as well as exoplanets near the breaking point ($\tirr=T_b=1790$ K).

\bibliographystyle{aasjournalv7}
\bibliography{rocky_IR_excess}

\end{CJK*}
\end{document}